\documentclass[%
 reprint,
 amsmath,amssymb,
 aps, showkeys,
]{revtex4-1}

\usepackage{graphicx}
\usepackage{dcolumn}
\usepackage{bm}
\usepackage{hyperref}
\usepackage{lipsum}
\usepackage{subfigure}
\usepackage{amsmath}

\graphicspath{ {./images/} }

\hypersetup{
    colorlinks = true,
    linkcolor = blue,
    citecolor = blue,
    urlcolor = blue,
}

\begin{document}

\preprint{APS/123-QED}

\title{Physical insights from the aspect ratio dependence of turbulence in negative triangularity plasmas}

\author{A. Balestri$^{1,*}$, J. Ball$^1$, S. Coda$^1$, D. J. Cruz-Zabala$^{2,3}$, M. Garcia-Munoz$^{2,3}$ and E. Viezzer$^{2,3}$}

\affiliation{\vspace{10pt}$^1$Ecole Polytechnique Fédérale de Lausanne (EPFL), Swiss Plasma Center (SPC), 1015 Lausanne, Switzerland\\
$^2$Dept. of Atomic, Molecular and Nuclear Physics, University of Seville, Seville, Spain\\
$^3$Centro Nacional de Aceleradores, (U. Seville, CSIC, J. de Andalusia), Seville, Spain}

\email{alessandro.balestri@epfl.ch}

\begin{abstract}
In this work, we study the impact of aspect ratio $A=R_0/r$ (the ratio of major radius $R_0$ to minor radius r) on the confinement benefits of Negative Triangularity (NT) plasma shaping. We use high-fidelity flux tube gyrokinetic GENE simulations and consider several different scenarios: four of them inspired by TCV experimental data, a scenario inspired by DIII-D experimental data and a scenario expected in the new SMART spherical tokamak. The present study reveals a distinct and non-trivial dependence. NT improves confinement at any value of $A$ for ITG turbulence, while for TEM turbulence confinement is improved only in the case of large and conventional aspect ratios. Additionally, through a detailed study of a large aspect ratio case with pure ITG drive, we develop an intuitive physical picture that explains the beneficial effect of NT at large and conventional aspect ratios. This picture does not hold in TEM-dominated regimes, where a complex synergistic effect of many factors is found. Finally, we performed the first linear gyrokinetic simulations of SMART, finding that both NT and PT scenarios are dominated by micro-tearing-mode (MTM) turbulence and that NT is more susceptible to MTMs at tight aspect ratio. However, a regime where ITG dominates can be found in SMART, and in this regime NT is more linearly stable.
\end{abstract}

\maketitle


\section{Introduction}\label{1}

The discovery of the High confinement mode (H-mode) in ASDEX \cite{Hmode_1989} opened the gates to a new era of research in the field of magnetically confined fusion plasmas. After exceeding a certain threshold of heating power, a tokamak plasma spontaneously creates a transport barrier at the edge, thereby enabling the formation of a region of steep pressure gradient. This is the so-called pedestal which greatly increases the temperature and density in the core of the plasma. The performance of H-mode plasmas is substantially better than standard Low confinement mode (L-mode) plasmas, which do not display a noticeable pedestal. However, the promise of H-mode is tempered by the appearance of powerful instabilities called Edge Localized Modes (ELMs) \cite{ELMs}, which are quasi-periodic relaxations of the pedestal that release large amounts of particles and energy into the plasma-facing components of the tokamak. In addition, the power released by an ELM is directly proportional to the power injected in the plasma, making ELMs extremely dangerous for the integrity of a power plant. A single ELM could seriously damage the first wall of the machine and compromise the steady production of energy. 

For these reasons, the European roadmap towards DEMO (DEMOnstration power plant) identified the ELMs as one of the most important issues to be solved \cite{DEMO} and recommends searching for methods to reduce the size of ELMs or for scenarios where they can be avoided without compromising confinement. A solution may come from plasma shaping.

Decades of plasma physics research have proven that giving a tokamak plasma a non-circular poloidal cross-sectional shape can greatly modify the stability of the plasma and its confinement of particles and energy. The poloidal cross-sectional shape of a plasma can be modified by changing the two most important shaping factors: elongation
\begin{equation}
    \kappa=\frac{Z_{max}-Z_{min}}{2a},
    \label{elongation}
\end{equation}
and triangularity
\begin{equation}
    \delta=\frac{2R_{geo}-(R_{top}+R_{bot})}{2a},
    \label{triangularity}
\end{equation}
where $Z_{max}$ and $Z_{min}$ are the maximum and minimum vertical positions of the flux surface, $R_{top}$ and $R_{bot}$ are the radial locations of the uppermost and lowermost points, $R_{geo}$ is the geometric center of the flux surface and $a$ is the minor radius.

The configuration that has become the standard scenario in the majority of modern tokamaks is the elongated \textit{dee-shaped} one, also called Positive Triangularity (PT) because the value of $\delta$ is larger than zero. The term \textit{dee} comes from the resemblance of the cross-section to the letter "D", when shown to the right of the tokamak axis of symmetry (as is typical). PT plasmas became popular because they have good MHD stability and can enter H-mode with less heating power. Of course, the downside is the aforementioned ELMs, the seriousness of which has been widely appreciated only more recently. An alternative that has gained interest is Negative Triangularity (NT), where the plasma is shaped such that $\delta$ is less than zero. A large number of experiments on several medium-size tokamaks (on TCV for several decades \cite{Pochelon,9,Coda_2022}, followed more recently by DIII-D \cite{6,7} and AUG \cite{ASDEX}) as well as many numerical studies \cite{4,merlo_jenko_2023,Marinoni_2009}, find that NT reduces turbulent transport, and can inhibit the transition to H-mode. These features allow a NT plasma to have an L-mode-like edge pressure profile and avoid ELMs, but with sufficiently increased pressure gradient at the edge and in the outer core to recover H-mode-like values of density and temperature in the core.

Despite the great number of studies on the topic, the physical processes underlying the beneficial effect of NT on turbulent transport are not fully understood yet. Moreover, as observed by Merlo and Jenko in \cite{merlo_jenko_2023}, it is key to study how other plasma parameters influence the impact of NT on confinement so we can use this knowledge to optimize. In this work, we will study the dependence of transport on the aspect ratio $A=R_0/r$ (the ratio of major radius $R_0$ to minor radius $r$) in NT plasmas. The motivation for such a study is twofold. First, studying how a NT plasma behaves at low $A$ gives insight into the feasibility of a NT Spherical Tokamak (ST), i.e. a tokamak characterized by a small aspect ratio. Second, studying the large aspect ratio limit greatly simplifies the physics and therefore the analysis. Moreover, as we will see, the conclusions will turn out to be valid also at the aspect ratios typical for a standard tokamak. As the aspect ratio is not a parameter that can be easily changed and scanned in real experiments, we make use of high-fidelity flux tube gyrokinetic GENE \cite{GENE} simulations, which allow us to explore the parameter space and capture the physical processes controlling turbulence. 

In this paper, we will see that a non-trivial picture will emerge from the aspect ratio scan, showing that NT has better confinement than PT at large and standard aspect ratio values, regardless of the nature of the turbulent regime. However, at tight $A$ NT becomes detrimental in Trapped Electron Mode (TEM) dominated regimes, while staying beneficial in Ion Temperature Gradient (ITG) mode dominated regimes. This complex dependency on the aspect ratio triggered a deeper study to understand the physical processes underlying the stabilization/destabilization of certain types of turbulence by a NT geometry. While an explanation of the behaviour at small $A$ is still missing, we will present a simple but compelling physical picture of why NT weakens ITG instability at large and standard aspect ratios. We refer to this understanding as simple because it relies only on two quantities, i.e. magnetic drifts and Finite Larmor Radius (FLR) effects, which in turn are easy to calculate from the shape of the plasma. Motivated by similarities between our model and the work done in \cite{Marinoni_2009} for TEM turbulence, we also tried to apply our model to a scenario with pure TEM drive, finding that FLR effects and magnetic drifts can only explain part of the physics for this kind of scenario.

Finally, to achieve a better understanding of the small $A$ regime and also use more realistic equilibria, we will consider the ST SMART \cite{SMART1,SMART2,MANCINI2023113833,AGREDANOTORRES2021112683}. This is a ST that only very recently started operation at the University of Seville, characterized by strong geometric flexibility, which should enable the creation of single-null and double-null NT and PT configurations. The input parameters for GENE simulations are taken from predictive simulations for SMART made with TRANSP \cite{Diego}. We will perform only local linear simulations. They will reveal which kind of turbulence dominates and how it responds to a change in various parameters. We will see that, because of the high $\beta$ typical of STs, large electron temperature gradient and large magnetic shear, turbulence is dominated by Micro-Tearing Modes (MTM). Moreover, we will find that NT is more sensitive to MTMs, with growth rates larger than those computed for PT. However, we will see that it should be possible to operate SMART in an ITG-dominated regime, where it is possible to exploit the beneficial effect of NT (consistently with previous simulations). We were not able to perform reliable NL simulations because of the challenges commonly encountered when simulating scenarios mainly driven by MTMs (e.g. lack of converged heat fluxes when using standard resolutions).

The remainder of this paper is organized as follows. In section \ref{2}, we describe briefly how GENE works and how it has been used to carry out the present study. In section \ref{3}, we present how the aspect ratio and turbulent regime can modify the beneficial effect of NT on confinement. In section \ref{6}, we present the first linear simulations performed for the SMART tokamak, providing predictions for the turbulent regime and the performance of the NT option with respect to PT. In section \ref{4}, we motivate and propose a simple physical picture of why NT weakens ITG turbulence in large and standard aspect ratio tokamaks. In section \ref{5}, we apply the model described in the previous section to a scenario with pure TEM drive. Finally, section \ref{7} draws some conclusions and provides a final discussion.

\section{Numerical model and methodology}\label{2}

This work consists exclusively of numerical simulations performed with the flux tube gradient-driven version of the GENE code \cite{GENE}. GENE is a physically comprehensive Eulerian gyrokinetic code that solves the Vlasov-Maxwell equations discretized on a 5-dimensional (5D) grid. This leads to a large set of coupled ordinary differential equations that can be solved as an initial value problem or as an eigenvalue problem. The code can be run in a linear mode if the nonlinear terms of the GK equation are neglected, or in a nonlinear mode when the nonlinear terms are retained. The former is considerably faster and is especially useful to quickly gain an understanding of the nature of turbulence. The latter is computationally more expensive but needed to physically model micro-turbulence. The real space is parametrized by a 3D system of non-orthogonal field-aligned coordinates $(x,y,z)$, which correspond respectively to the radial, binormal, and parallel (to the magnetic field \textbf{B}) coordinates. They are defined by
\begin{equation}
    \begin{cases}
        x=x(\psi)\\
        y=C_y\left(q(\psi)\theta-\varphi\right),\\
        z=\theta
    \end{cases}
    \label{coordinates}
\end{equation}
where $\psi$ is the poloidal magnetic flux, $\varphi$ the toroidal angle, $\theta$ the straight field-line poloidal angle, $q$ is the safety factor and $C_y$ is a constant. The remaining two coordinates of the 5D discretized grid, $v_\parallel$ and $\mu$, describe velocity space and correspond to the parallel velocity and the magnetic moment (defined as $\mu=(m_\sigma v_\perp^2)/(2B)$, where $m_\sigma$ is the mass of the spiecies $\sigma$ and $v_\perp$ the perpendicular velocity). In a flux tube, the background values of density, temperature, and flow (as well as their gradients) are evaluated at the flux surface of interest and kept constant across the simulation domain. This choice is justified if the radial scale of turbulence is much smaller than the radial scale of variation of equilibrium quantities (machine scale), which is fairly well satisfied for TCV and better satisfied in larger devices like DIII-D or reactor-scale devices. The real space simulation domain corresponds to a small rectangle extended in $x$ and $y$, that follows the magnetic field lines along $z$. Periodic boundary conditions are used for the perpendicular coordinates $x$ and $y$, while pseudo-periodic boundary conditions are applied to $z$. For more details, the interested reader is referred to \cite{40}. 

To carry out a comprehensive analysis, we will consider scenarios characterized by different turbulent regimes, i.e. regimes with different dominant turbulent drive. Moreover, to maintain realism, we will use discharges produced in TCV and DIII-D, and scenarios predicted for SMART. In total, we will consider six different cases: a realistic TEM dominated scenario (rTEM), a realistic ITG dominated scenario (rITG), a pure density gradient-driven TEM scenario (pTEM), two pure ion temperature gradient-driven scenarios (pITG-1 and pITG-2), and an MTM dominated scenario (the one relative to SMART).

GENE can use different methods or codes to reconstruct the magnetic equilibrium at a specific flux surface. These can be divided into numerical reconstruction codes (CHEASE \cite{CHEASE}, gist, and tracer-efit) and analytical models ($s-\alpha$ and Miller). For all the simulations presented in this work, we used the local equilibrium Miller model, because, as an analytical model, it allows us to easily change the geometry by changing a few parameters. As shown in \cite{Candy_2009}, a Miller equilibrium is completely defined by a total of 14 scalar quantities. As shown in equation \ref{eq_Miller}, the parametrization of the flux surface shape requires 6 parameters: the geometric axis $R_0$, the elevation $Z_0$, the aspect ratio $A=R_0/r$, the elongation $\kappa$, the triangularity $\delta$ and the squareness $\zeta$.
\begin{equation}
    \begin{cases}
        R(\theta)=R_0[1+A^{-1}\cos{\left(\theta+\arcsin{(\delta\sin{\theta})}\right)}]\\
        Z(\theta)=Z_0+\kappa r\sin{\left(\theta+\zeta\sin{(2\theta)}\right)},
    \end{cases}
    \label{eq_Miller}
\end{equation}
where $R$ and $Z$ are the radial and axial cylindrical coordinates as a function of the poloidal angle $\theta$ ($\theta=0$ corresponds to the outboard midplane, while $\theta=\pm\pi$ the inboard midplane). The computation of the poloidal field necessitates 8 additional parameters: the elongation shear $s_\kappa$, the triangularity shear $s_\delta$, the squareness shear $s_\zeta$, the safety factor $q_0$, the magnetic shear $\hat{s}$, the pressure gradient $p'$, and the Shafranov Shift given by $\partial_rR_0$, and the variation of elevation $\partial_rZ_0$. 

To perform our simulations, we used Miller to model relevant flux surfaces in actual experimental magnetic equilibria. We flipped the triangularity (from NT to PT or vice versa) and changed the aspect ratio varying the minor radius. During these operations, the values of density and temperature, and their logarithmic gradients, were kept fixed. This procedure is commonly used in GK simulations and enables one to isolate the effect of geometry on transport. The numerical convergence of all simulations has been tested by resolution studies. More details will be given in the next sections, but, depending on the simulation, we used $L_x\sim[110-180]\rho_i$ and $L_y\sim[125-250]\rho_i$ for the radial and binormal widths of the simulation box with $[128-256]$ $k_x$ and $[32-64]$ $k_y$ modes. Depending on the aspect ratio, we used between 32 to 48 grid points in $z$. The size and resolution of velocity space have been kept fixed: $L_v=3v_{th,\sigma}$, $L_\mu=9\mu B_0/T_\sigma$, $n_v=42$ and $n_\mu=12$, where $v_{th,\sigma}=\sqrt{2T_\sigma/m_\sigma}$ is the thermal velocity, $B_0$ the toroidal magnetic field at $R_0$ on the flux surface of interest, $m_\sigma$ the reference mass and $T_\sigma$ the reference temperature.


\section{Aspect ratio scan}\label{3}

To carry out the most realistic analysis possible, we based our parameters on an Electron-Cyclotron-Resonance (ECR) heated TCV NT discharge (\#68145), a Neutral Beam Injection (NBI) heated TCV NT discharge (\#69682), and a high power NBI-heated DIII-D PT discharge (\#176492). We reconstructed the magnetic equilibria with Miller, choosing a radius of analysis (defined here as the normalized square root of the toroidal flux) $\rho_{tor}=0.8$ for TCV and $\rho_{tor}=0.75$ for DIII-D. Then we flipped the sign of $\delta$ and $s_\delta$ and we artificially changed the aspect ratio keeping fixed the major radius and changing the minor radius. At the chosen radial locations, the logarithmic gradients were computed from the experimental data and then were kept fixed while the geometry was changed. 

\begin{table}
    \centering
    \begin{tabular}{l|ccccc}
    \toprule
        &rTEM & pTEM & pITG-1 & pITG-2 & rITG \\ \hline
        $R/L_{Te}$&14.54 & 0 & 0 & 0 & 6.37\\ 
        $R/L_{Ti}$&8.48 & 0 & 8.48 & 12.89& 8.98\\ 
        $R/L_{ne}$&6.01 & 7.81 & 0 & 0 & 1.47\\
        $R/L_{nC}$&9.16 & - & - & - & 1.30\\
        $T_e$ [keV]&0.41 & 0.41 & 0.41 & 0.31 & 1.47\\ 
        $T_e/T_i$&0.98 & 0.98 & 0.98 & 1.37 & 1.17\\ 
        $n_e\,[10^{19}m^{-3}]$& 1.95 & 1.95 & 1.95 & 1.11 & 4.25\\ 
        $n_C/n_e$&0.013 & - & - & - & 0.018\\
        $\beta [\%]$ &0.16 & 0.16 & 0.16 & 0.07 & 0.71\\
        $\nu_{ei}$ [$s^{-1}$] &0.84 & 0.84 & 0.84 & 0.838 & 0.45\\
        \hline
        $A$& 5 & 5 & 5 & 5 & 3.7\\ 
        $q$&2.01 & 2.01 & 2.01 & 2.20 & 3.41\\ 
        $\hat{s}$&1.6 & 1.6 & 1.6 & 2.02 & 2.43\\
        $\kappa$ & 1.37 & 1.37 & 1.37 & 1.31 & 1.47\\
        $|\delta|$ &0.16 & 0.16 & 0.16 & 0.25 & 0.21\\
        $\zeta$ & -0.07 & -0.07 & -0.07 & 0.24 & -0.31\\
        $s_\kappa$ & 0.19 & 0.19 & 0.19 & 0.16 & 0.35\\
        $|s_\delta|$& 0.33 & 0.33 & 0.33 & 0.59 & 0.54\\
        $s_\zeta$ &-0.21 & -0.21 & -0.21 & 0.12 & -0.14\\\toprule
    \end{tabular}
    \caption{\label{input} Logarithmic gradients of electron temperature $R/L_{Te}$, ion temperature $R/L_{Ti}$, electron density $R/L_{ne}$, carbon density $R/L_{nC}$, as well as the ion-electron $T_e/T_i$ temperature ratio, electron density $n_e$, the carbon to electron density $n_C/n_e$, the local electron $\beta$, the electron-ion collisional frequency $\nu_{ei}$ and geometric parameters needed to specify the Miller equilibrium.}
\end{table}

\begin{figure}[]
    \centering
    \includegraphics[width=\linewidth]{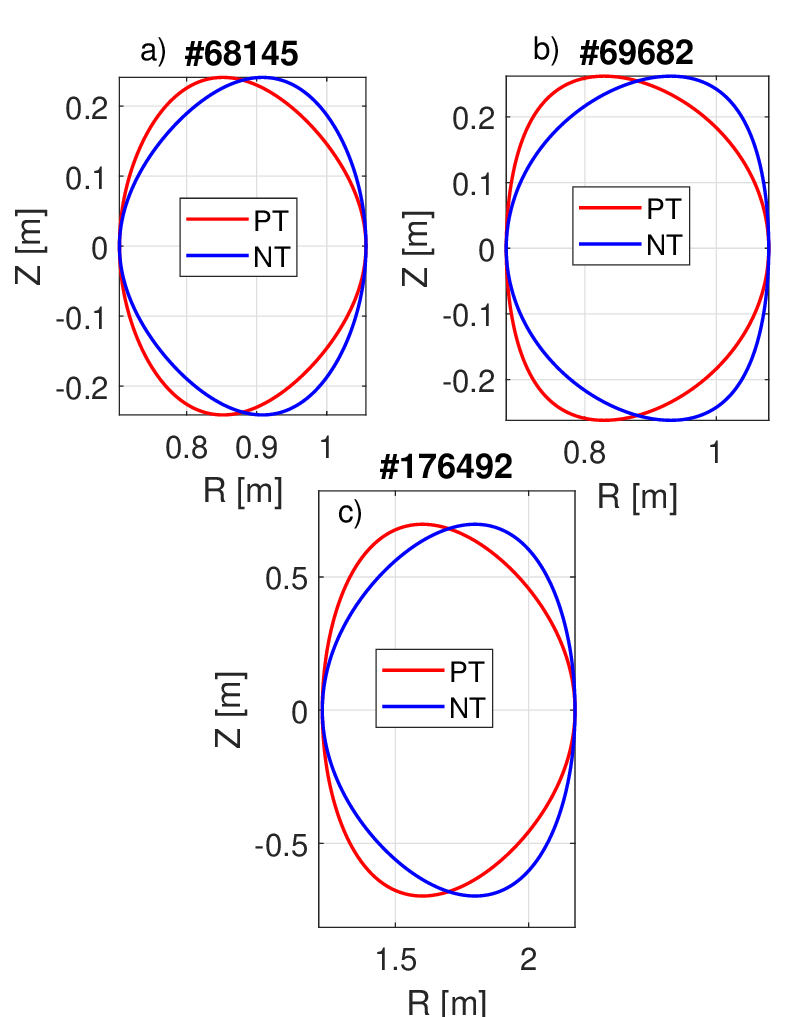}
    \caption{The PT and NT flux surface shapes of the simulated scenarios: (a) TCV \#68145 at $\rho_{tor}=0.8$, (b) TCV \#69682 at $\rho_{tor}=0.8$,  (c) DIII-D \#176492 at $\rho_{tor}=0.75$}
    \label{shapes}
\end{figure}

\begin{figure*}
    \centering
    \includegraphics[width=\textwidth]{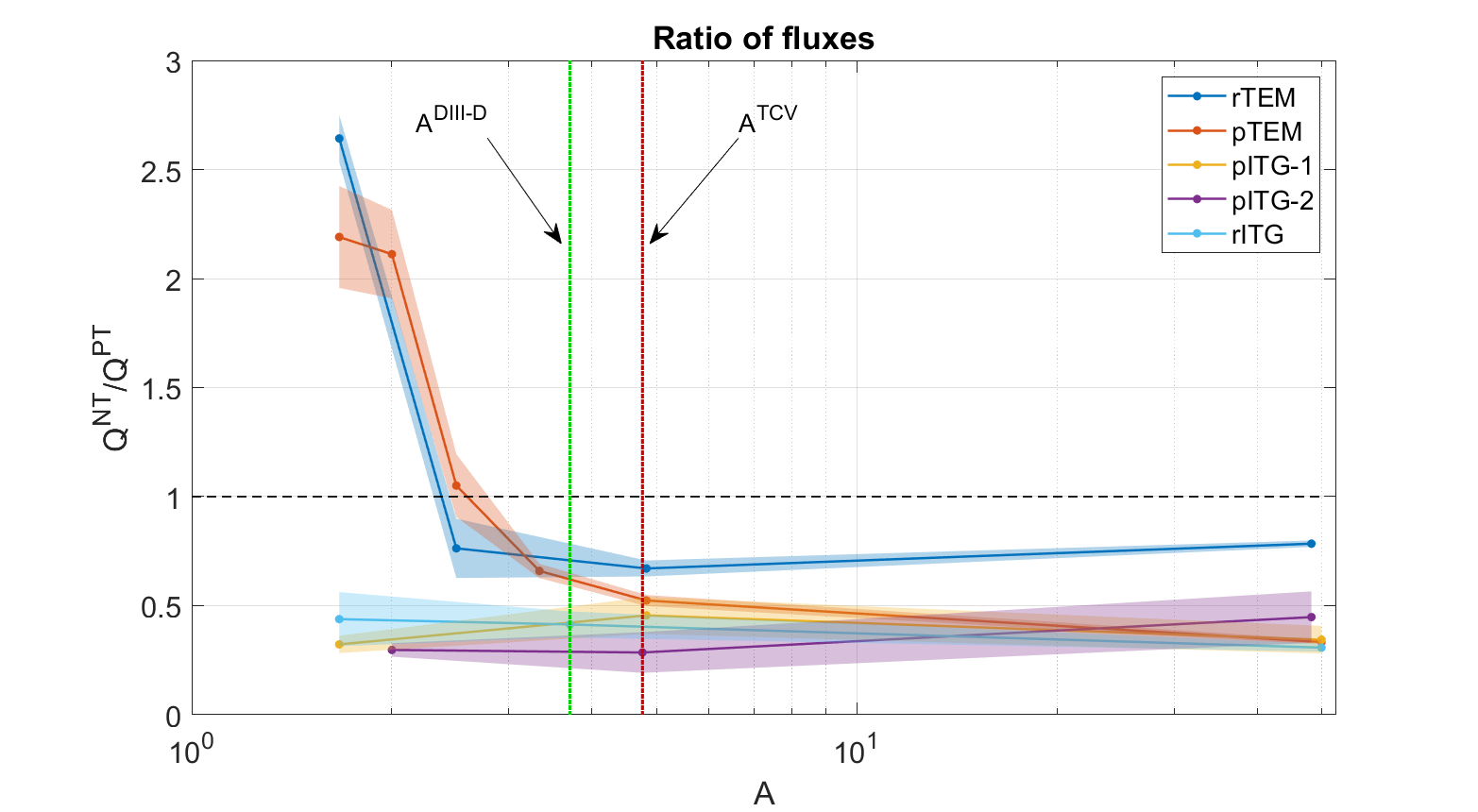}
    \caption{The heat flux in the NT geometry divided by that of the PT geometry as a function of aspect ratio. The red and green dotted lines correspond to the real aspect ratio of the considered flux surface in TCV and DIII-D respectively. Spherical tokamaks are shown on the left of the plot.}
    \label{Qratio_A}
\end{figure*}

\begin{figure*}
    \centering
    \includegraphics[width=\textwidth]{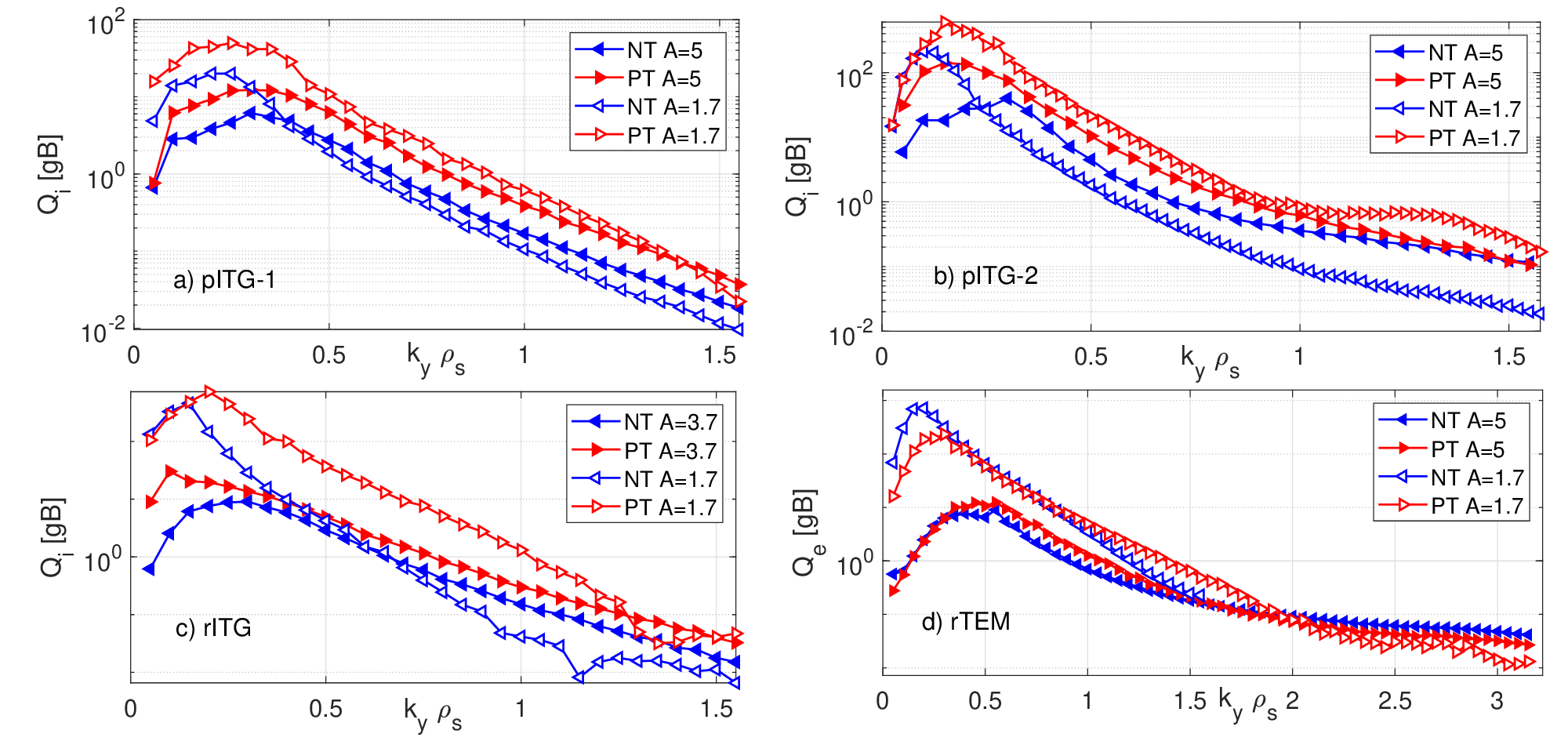}
    \caption{Heat flux spectra for different turbulent regimes, triangularities, and aspect ratios. Results for NT are in blue, PT in red, conventional $A$ with full triangles, and small $A$ with empty triangles.}
    \label{spectra}
\end{figure*}

\begin{figure*}
    \centering
    \includegraphics[width=\textwidth]{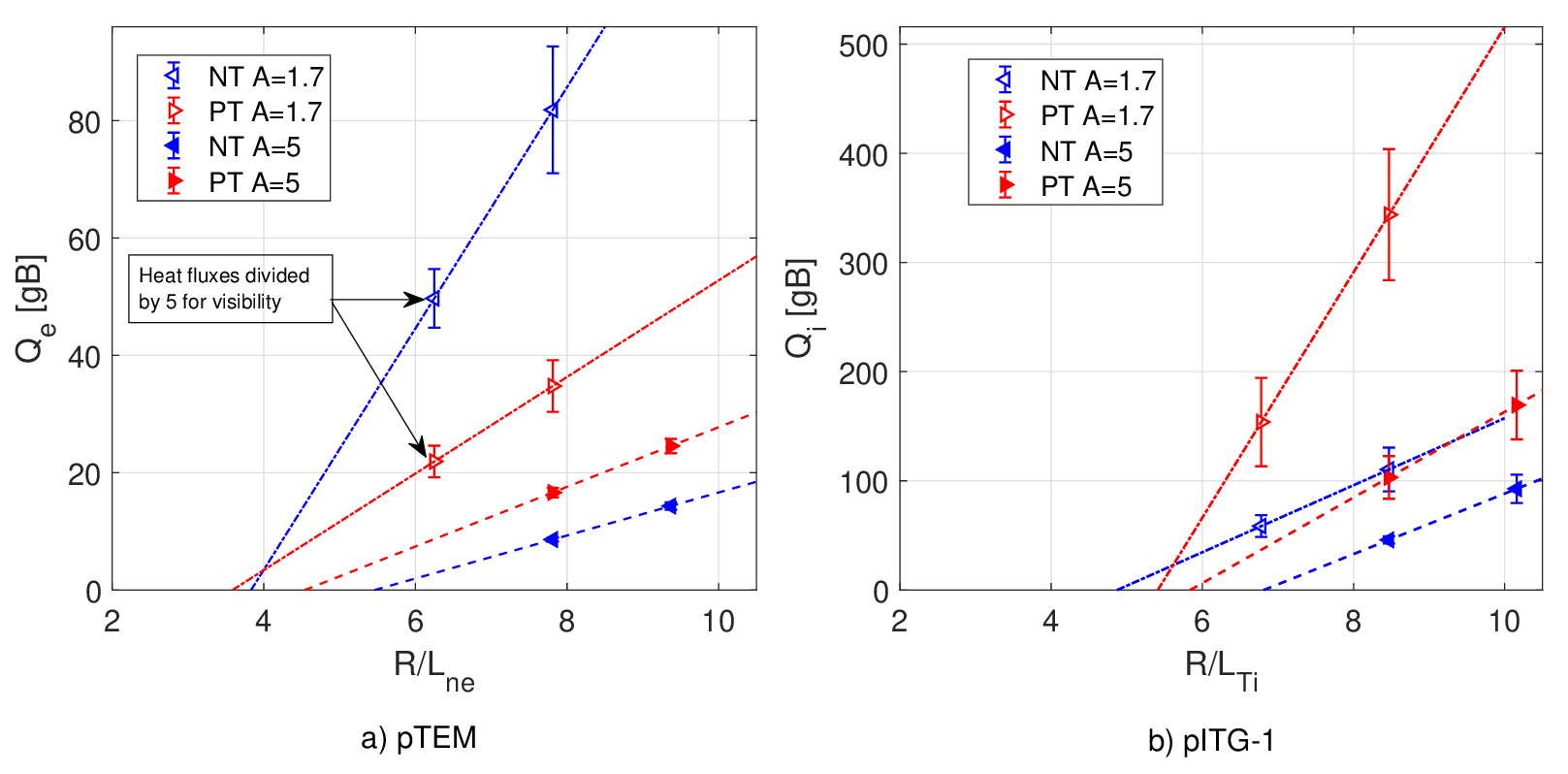}
    \caption{Sensitivity of the heat fluxes to the main driving gradient of turbulence for two different aspect ratios, in the (a) pTEM case and (b) the pITG-1 case. For all the pTEM cases with $A=1.7$, the heat fluxes have been divided by 5 to clearly plot data. The error bars correspond to the standard deviation of the average value computed from the time traces of the heat fluxes.}
    \label{stiffness}
\end{figure*}

We chose discharges \#68145 and \#176492 to have cases dominated by TEM and ITG turbulence respectively. Originally, \#69682 was chosen to provide an ITG-dominated case, but linear GK simulations revealed that it was TEM-dominated. Additionally, we modified the logarithmic gradients of \#68145 and \#69682 to create idealized cases, namely two pure ion temperature gradient-driven scenarios (pITG-1 and pITG-2) and a pure density gradient-driven TEM one (pTEM). Some of the most important input parameters used in the simulations are reported in table \ref{input}, while the shapes of the three original scenarios with their flipped counterparts are shown in figure \ref{shapes}. For the realistic TCV and DIII-D simulations collisions, electromagnetic effects and carbon as the main impurity were retained. The pITG and pTEM simulations neglected impurities, but collisions and electromagnetic effects were kept. All simulations used a kinetic treatement for the electrons.

We started by performing linear simulations for each scenario to establish the turbulence drives stated above. These results are presented in appendix \ref{A}. We proceeded with nonlinear simulations where we scanned the aspect ratio by varying the minor radius of the flux surface, while keeping the major radius fixed. 
Figure \ref{Qratio_A} shows the ratio between the nonlinear heat fluxes from the NT cases ($Q^{NT}$) and the PT cases ($Q^{PT}$). When the ratio of fluxes is below one, the NT configuration has better confinement, while the opposite is true when the ratio is above one. The shaded areas correspond to the standard deviation of the averaged values computed from the time traces of the heat fluxes. 

Figure \ref{Qratio_A} clearly shows that the influence of NT on transport has a complex and non-trivial dependency on aspect ratio. Regardless of the turbulent regime, at large and conventional values of A, NT always has better confinement than PT. In this range of aspect ratio values, the reduction of the heat flux by NT varies between 1.5 for the TEM-dominated regime to a factor of 2.5 for the pITG scenarios. For STs (i.e. $A<2.5$), we find a bifurcation. When turbulence is purely ITG or ITG-dominated, NT still greatly stabilizes turbulence, reducing the heat flux by a factor of $\sim 2$. Instead, for a pure-TEM or TEM-dominated regime, NT becomes detrimental, increasing the heat flux by a factor of $\sim 2.5$.

Heat flux spectra are plotted in figure \ref{spectra} for different regimes using two values of aspect ratio: conventional and spherical. We are not showing the spectra for the large aspect ratio cases because they are similar to conventional A. We show only the most important component, i.e. electron or ion channel, depending on the considered regime. From figures \ref{spectra}(a), \ref{spectra}(b) and \ref{spectra}(c), which correspond to ITG regimes, we can see that at conventional $A$ the reduction of heat flux is equally distributed among all the considered $k_y$ modes. Instead, when we move to a smaller $A$, the reduction is more pronounced for modes with $k_y\rho_s$ bigger than 0.2. If we now look at figure \ref{spectra}(d), which corresponds to the realistic TEM case, we can see that the detrimental effect of NT comes from small $k_y$ modes at tight A. These are the modes that are most strongly destabilized when one decreases the aspect ratio for TEM turbulence.

To achieve a deeper insight, it is important to study the stiffness of transport, i.e. how the heat fluxes change when the value of the driving gradient is changed. Since the driving gradient is easiest to identify for pITG and pTEM, we have performed simulations for these cases with increased or decreased $R/L_{Ti}$ and $R/L_{ne}$, respectively. The study of the stiffness is extremely important to make PT/NT comparisons at fixed heat flux (i.e. gradient driven simulations).

Figure \ref{stiffness} displays the results. At conventional aspect ratio, flipping the triangularity does not change the stiffness but rather increases the critical gradient, i.e. the gradient at which modes become unstable. This results in reduced turbulent transport and better confinement properties. It also means that, at fixed heat flux, one can produce a plasma with increased gradients. The observation that, in a standard $A$ tokamak, NT primarily affects the critical gradient and not the stiffness is in agreement with many other works \cite{5,merlo_jenko_2023}. In contrast, at small $A$ we can see the opposite. Indeed, flipping the triangularity does not alter the critical gradient much but does change the stiffness. In the pITG regime NT reduces the stiffness, while in the pTEM regime NT increases it. All these observations indicate that turbulence responds to a change in shape very differently in conventional tokamaks and STs. 

\section{SMART}\label{6}

In this section, we present the first gyrokinetic simulations of the PT and NT scenarios foreseen for SMART.

SMART is a novel, compact, spherical tokamak ($1.4\leq A\leq3.0$) that has recently begun operations at the University of Seville. SMART has a strong geometric flexibility that allows it to produce plasmas with an elongation of $\kappa\leq3.0$ and a triangularity between $-0.6\leq\delta\leq0.6$. Both single-null and double-null configurations can be produced. For more technical details, the interested reader is referred to \cite{SMART1,SMART2,MANCINI2023113833,AGREDANOTORRES2021112683}.

\begin{figure}[]
    \centering
    \includegraphics[width=0.96\linewidth]{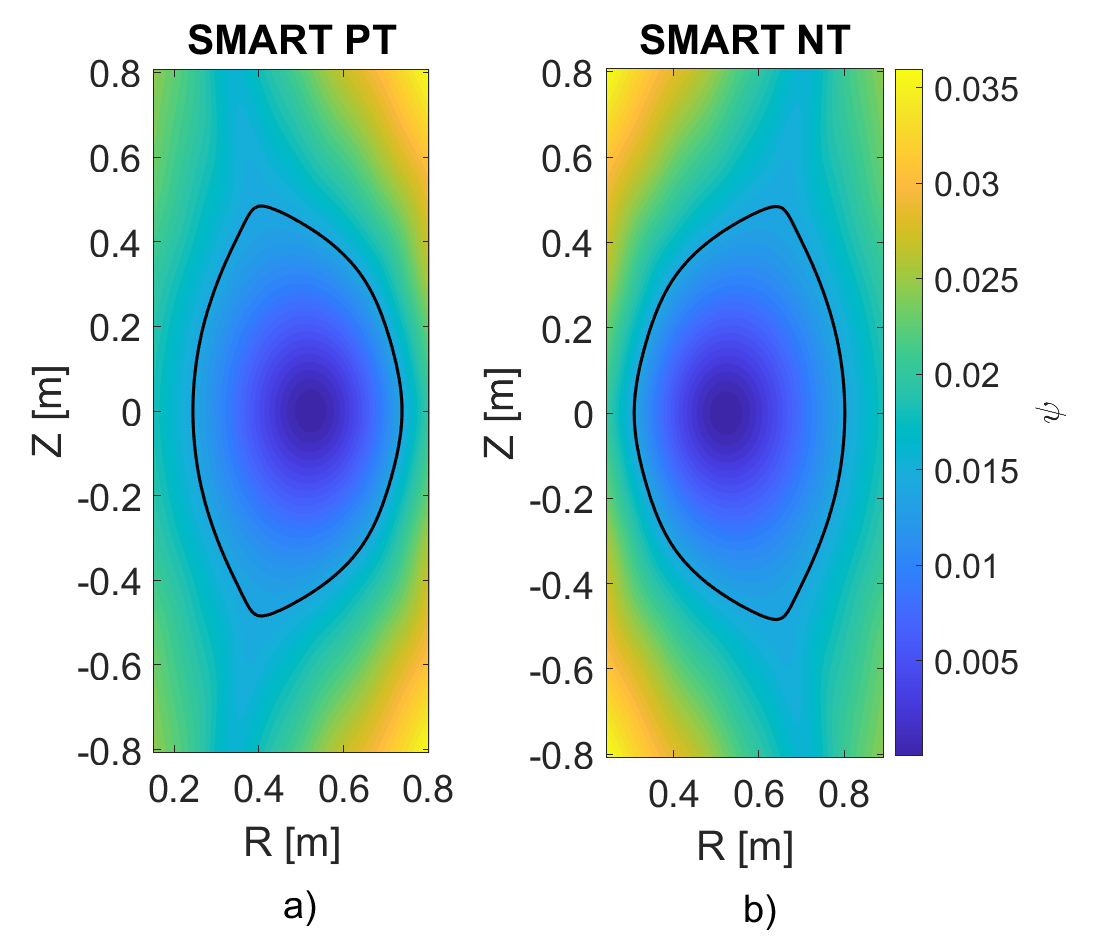}
    \caption{The poloidal flux $\psi$ predicted for the (a) PT and (b) NT options in double-null SMART discharges. The black solid line is the Last Closed Flux Surface (LCFS).}
    \label{SMART_eq}
\end{figure}

\begin{figure}[]
    \centering
    \includegraphics[width=\linewidth]{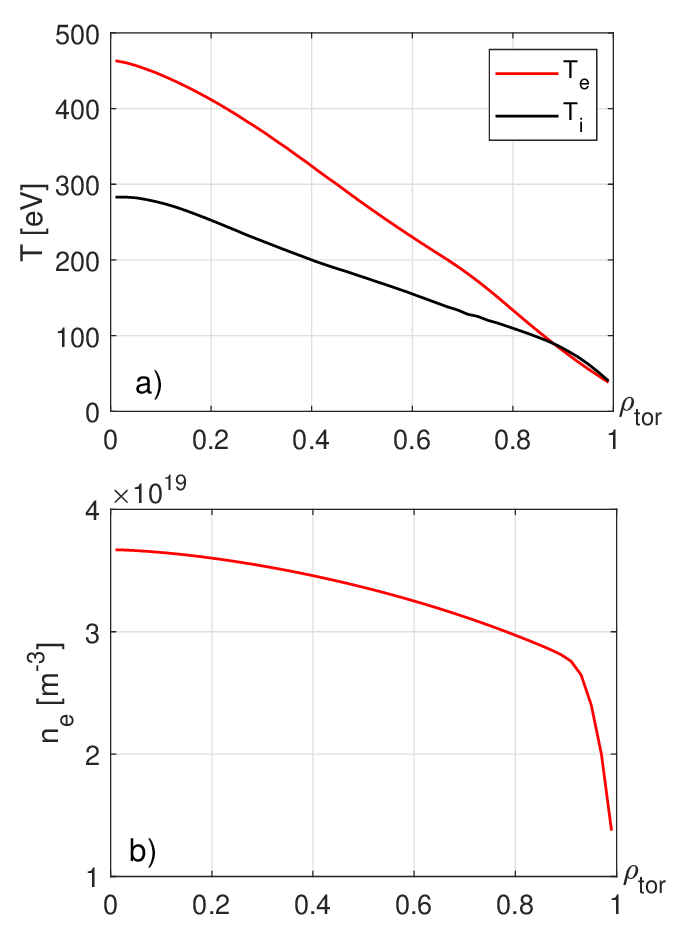}
    \caption{The radial profiles of (a) electron and ion temperature and (b) electron density predicted for the PT scenario in SMART.}
    \label{SMART_prof}
\end{figure}



\begin{table}[]
\begin{tabular}{c|c}\toprule
 $R/L_{Te}$ & 15.1 \\
 $R/L_{Ti}$ & 9.2 \\
 $R/L_{n}$ & 2.0 \\
 $T_e$ [keV] & 0.10\\
 $T_i/T_e$ & 0.99 \\
 $n_{e}\;[10^{19}m^{-3}]$ &  2.83\\\hline
 $\beta_e [\%]$ & 0.56 \\
 $\nu_{ei}$ & 8.62 \\
 \hline
 $A$& 2.2\\ 
 $q$&3.33\\ 
 $\hat{s}$&3.71 \\
 $\kappa$ & 1.76  \\
 $|\delta|$ & 0.16\\
 $\zeta$ & -0.04\\
 $s_\kappa$ & 0.66\\
 $|s_\delta|$& 0.96\\
 $s_\zeta$ &-0.28\\\toprule
\end{tabular}
\caption{\label{GK_par}GENE input parameters at $\rho_{tor}=0.88$}
\end{table}


\begin{figure}[]
    \centering
    \includegraphics[width=0.9\linewidth]{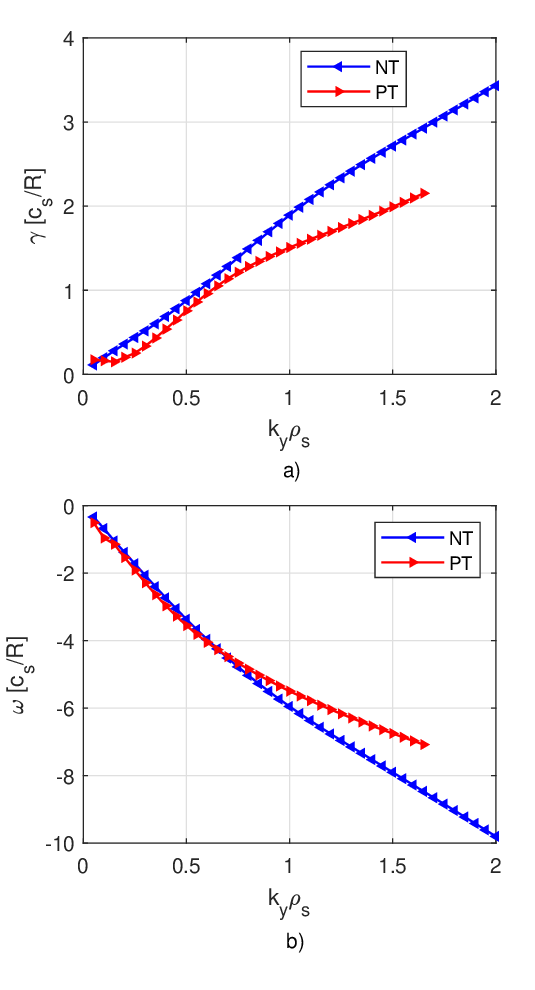}
    \caption{The (a) growth rate and (b) real frequency spectra of the most unstable linear modes in PT (red) and NT (blue) SMART equilibria.}
    \label{SMART_ky}
\end{figure}

\begin{figure}[]
    \centering
        \includegraphics[width=\linewidth]{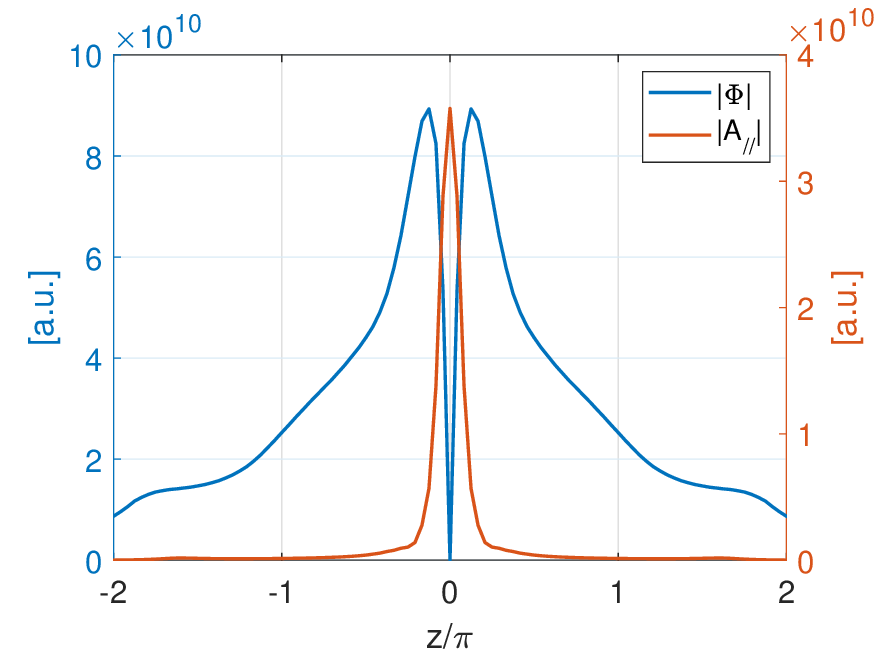}
    \caption{Modulus of the electrostatic potential (blue) and parallel component of the vector potential (orange) as a function of the poloidal angle.}
    \label{ball_0p4}
\end{figure}

Given the results of the previous section, experimental results from SMART will be very interesting. All the previous simulations are case studies and can partially predict how a NT ST will behave in the limit of certain idealized regimes. To more realistically model SMART, the input parameters for GK simulations have been obtained from TRANSP predictive simulations for the PT scenario \cite{Diego}. In figure \ref{SMART_eq} we show the magnetic equilibria and in figure \ref{SMART_prof} we show the kinetic profiles predicted by TRANSP for the PT scenario. This is an ohmic heated scenario and, since intrinsic rotation is hard to predict, rotation and flow shear were not considered in the following gyrokinetic simulations.

All linear simulations have been performed using Miller parametrization to construct a local magnetic equilibrium at $\rho_{tor}=0.88$. We chose this location because it is sufficiently close to the edge to have large $|\delta|$, but sufficiently far from the edge to avoid the region with steep gradients. The simulations retained kinetic electrons, collisions and electromagnetic effects. Other important parameters used in the simulations are listed in table \ref{GK_par}.  As before, to isolate the effect of the geometry, we kept these parameters fixed and flipped the shape. 
In this section, we will only show linear simulations as we were not able to achieve reliable NL simulations, which are subject to a future publication.

As already mentioned, one has to be careful in assessing the performance of a scenario using only linear simulations. Indeed, the physics of nonlinear saturation is missing and we cannot compute the final heat fluxes without it. However, a linear analysis can allow one to determine the dominant types of instability as well as to study the linear stability of the modes. We started with a $k_y$ scan (at $k_x=0$) to determine the type of turbulence that dominates at the ion scale, i.e. $k_y\rho_s=[0.05-2.0]$. To be consistent with the work shown in previous sections, we did not consider electron scale turbulence (i.e. $k_y\rho_s\gg2.0$). However, Electron Temperature Gradient (ETG) modes can play an important role in spherical tokamaks \cite{ETG1,ETG2}. This is left as future work. The results of the simulations at the ion scale are displayed in figure \ref{SMART_ky}. Comparing the magnitude of the growth rates, we see that NT has larger values than PT. This does not necessarily imply larger nonlinear heat fluxes, but does mean that the modes grow faster. However, the most interesting observation is the nature of turbulence. For the entire ion scale, the spectrum is dominated by MTMs. Indeed, looking at the electrostatic potential $\phi$ and the parallel vector potential $A_\parallel$ in figure \ref{ball_0p4}, we can see that these modes have tearing parity.

It is well-known that STs can be dominated by MTM turbulence \cite{Applegate_2007,MASTU}. MTM turbulence was not significant in the previous sections, as $\beta$ and the ratio of electron to ion temperature gradient were smaller. Therefore, to better characterize MTM turbulence in STs and how it behaves depending on triangularity, we performed parameter scans for the $k_y\rho_i=0.4$ mode.

\begin{figure}[]
    \centering
    \includegraphics[width=\linewidth]{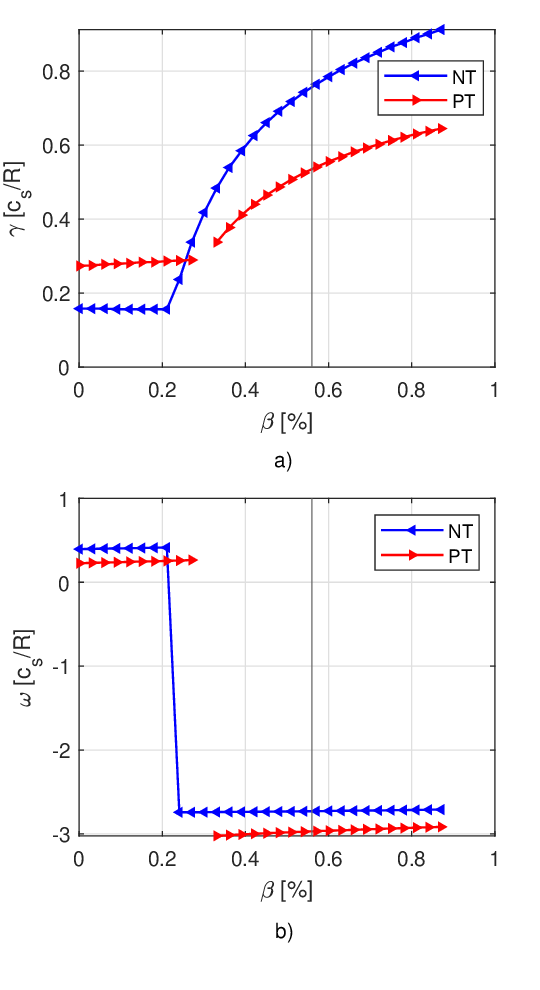}
    \caption{The (a) growth rate and (b) real frequency of the $k_y\rho_s=0.4$ mode as a function of the electron $\beta$ in PT (red) and NT (blue) SMART equilibria.}
    \label{SMART_beta}
\end{figure}


Figure \ref{SMART_beta} shows the results of a $\beta$ scan. First, we can identify two distinct regions depending on the value of $\beta$. For $\beta<0.2\%$, both PT and NT are dominated by ITG turbulence. It is interesting to notice that we recover the same result that we observed for the TCV and DIII-D low $A$ cases: NT has ITG modes with smaller growth rates. At larger values of $\beta$ we can identify a transition to the MTM regime. Importantly, NT's transition to MTM-dominated regime occurs at $\beta\sim0.2\%$, while PT's transition is at $\beta\sim0.3\%$. This suggests that NT is more sensitive to electromagnetic instabilities. At even larger $\beta$, we can see that MTMs are always stronger in NT than PT. The reason for this behaviour is still unclear and is a topic for future work.

\begin{figure}[]
    \centering
    \includegraphics[width=\linewidth]{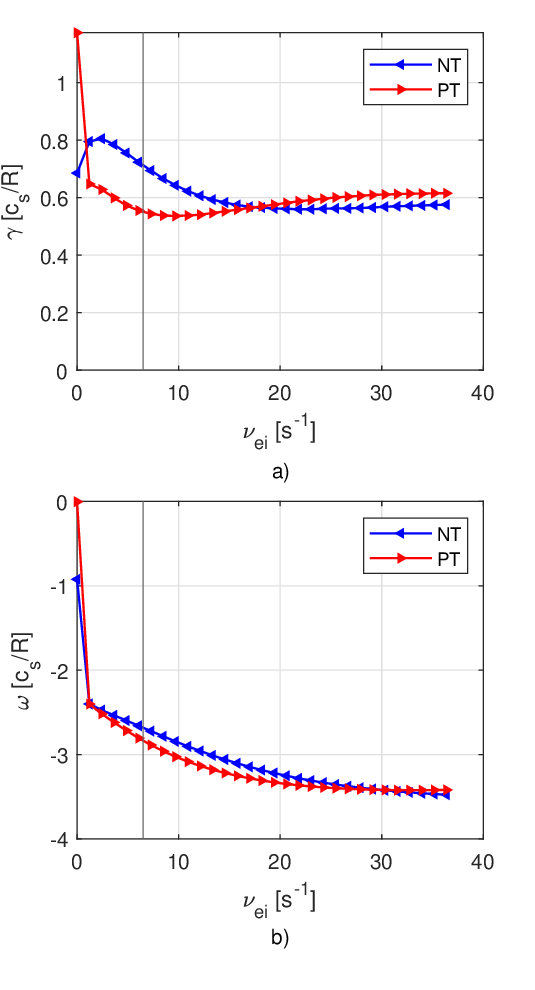}
    \caption{The (a) growth rate and (b) real frequency of the $k_y\rho_s=0.4$ mode as a function of the electron-to-ion collisionality $\nu_{ei}$ in PT (red) and NT (blue) SMART equilibria.}
    \label{SMART_coll}
\end{figure}


\begin{figure}[]
    \centering
    \includegraphics[width=\linewidth]{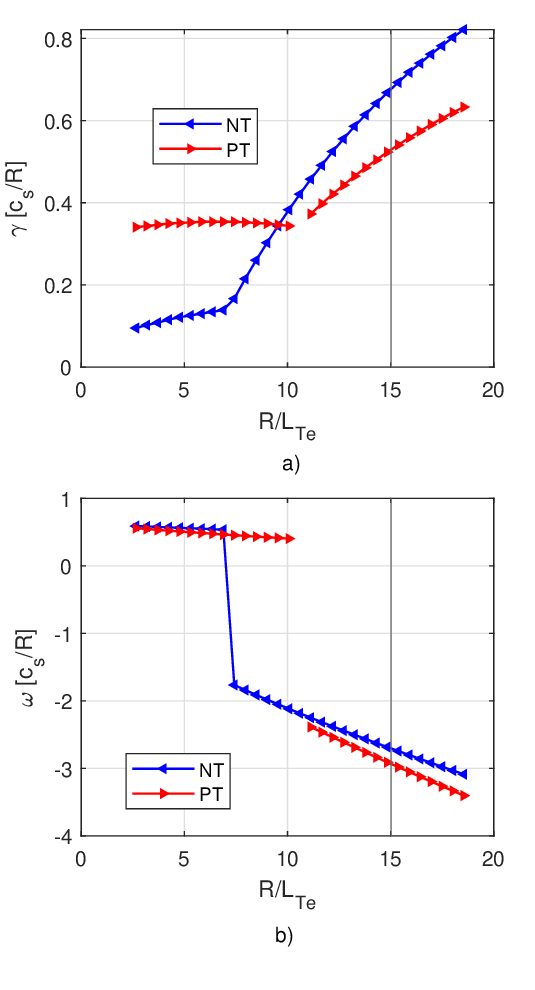}
    \caption{The (a) growth rate and (b) real frequency of the $k_y\rho_s=0.4$ mode as a function of the electron temperature gradient $R/L_{Te}$ in PT (red) and NT (blue) SMART equilibria.}
    \label{SMART_omte}
\end{figure}


\begin{figure}[]
    \centering
    \includegraphics[width=\linewidth]{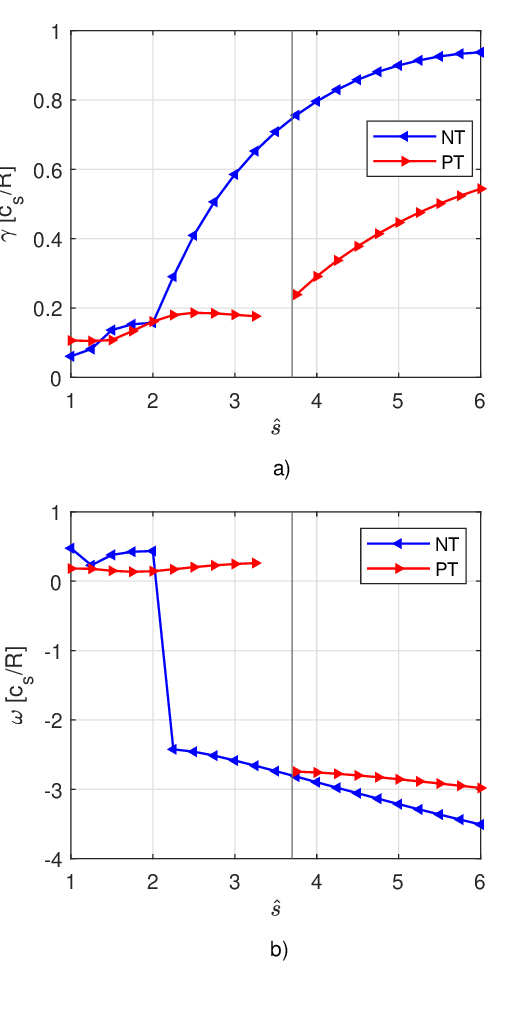}
    \caption{The (a) growth rate and (b) real frequency of the $k_y\rho_s=0.4$ mode as a function of the magnetic shear $\hat{s}$ in PT (red) and NT (blue) SMART equilibria.}
    \label{SMART_shear}
\end{figure}


\begin{figure*}[]
    \centering
    \begin{subfigure}[]
        { \includegraphics[width=0.45\linewidth]{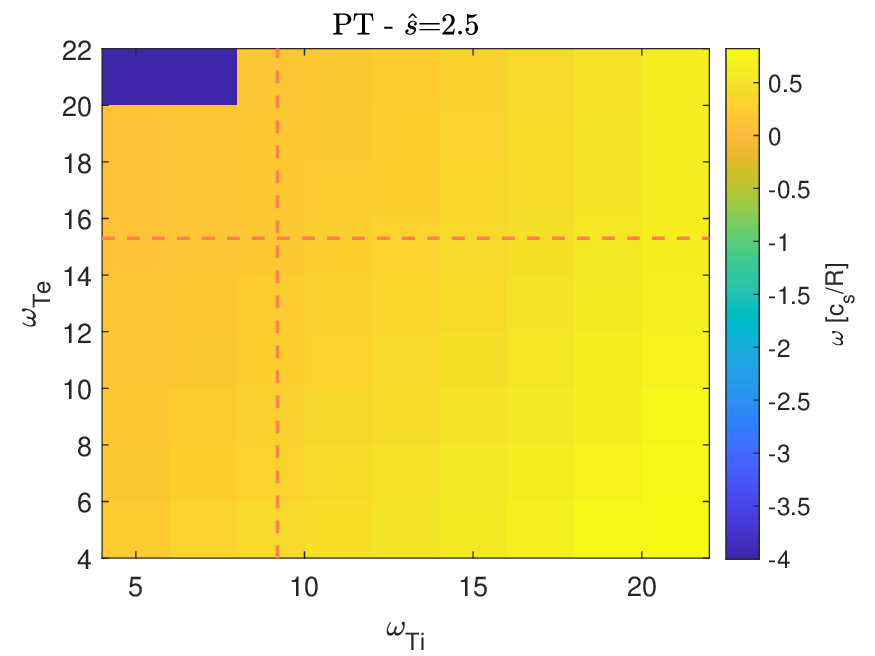}}
    \end{subfigure}
    \begin{subfigure}[]
        { \includegraphics[width=0.45\linewidth]{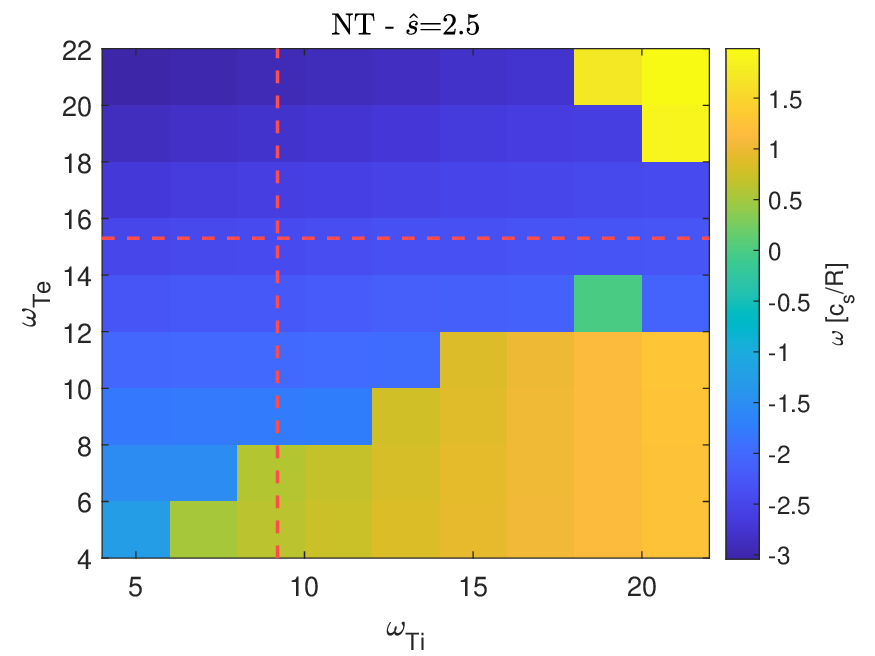}}
    \end{subfigure}
    \begin{subfigure}[]
        { \includegraphics[width=0.45\linewidth]{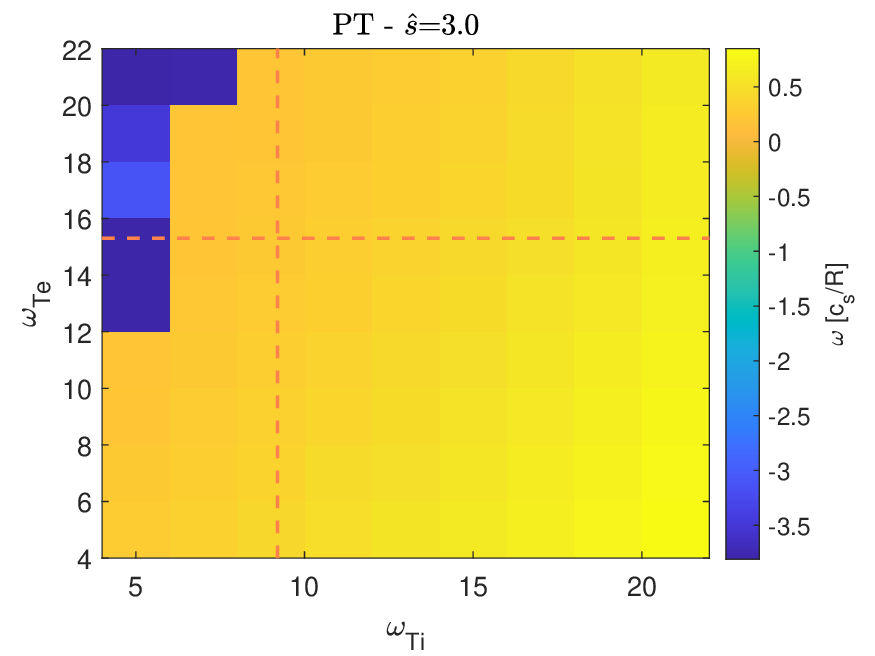}}
    \end{subfigure}
    \begin{subfigure}[]
        { \includegraphics[width=0.45\linewidth]{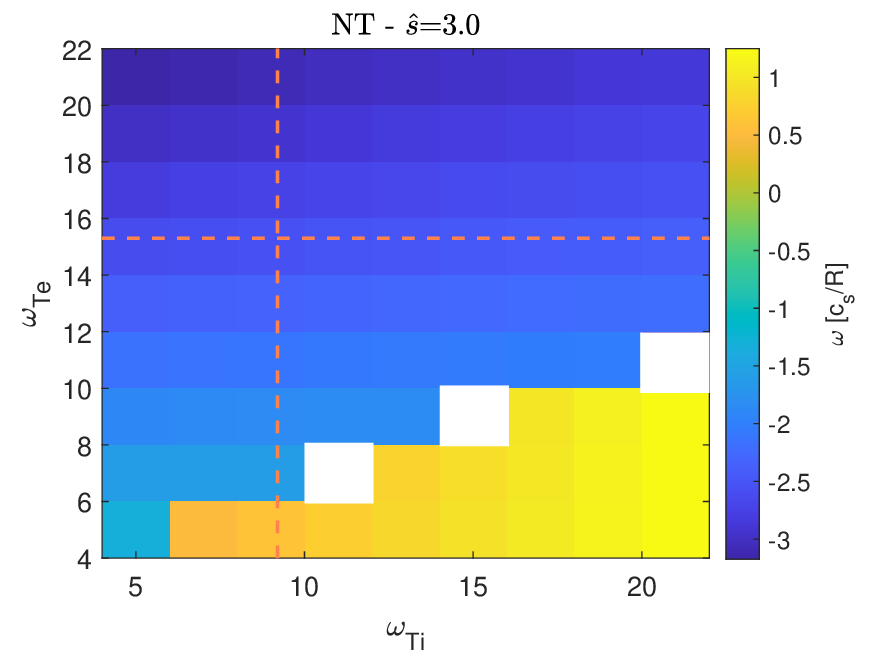}}
    \end{subfigure}
    \begin{subfigure}[]
        { \includegraphics[width=0.45\linewidth]{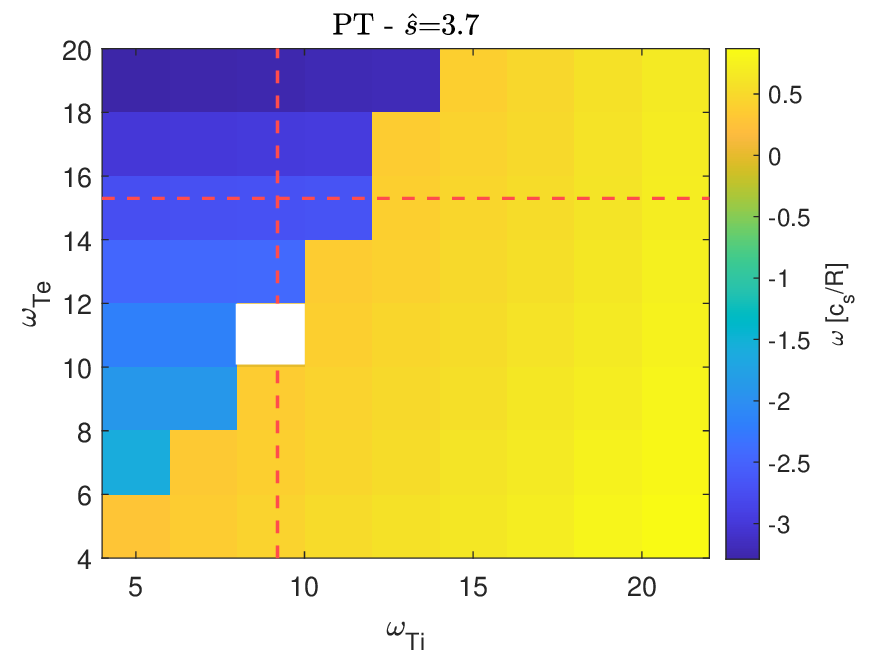}}
    \end{subfigure}
    \begin{subfigure}[]
        { \includegraphics[width=0.45\linewidth]{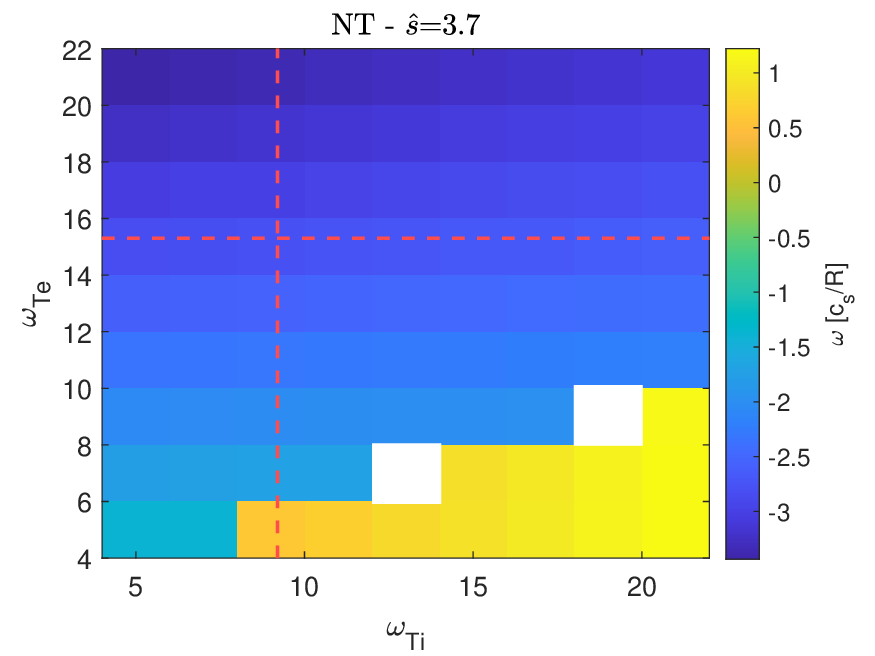}}
    \end{subfigure}
    \caption{Two-dimensional scans of ion and electron temperature gradients for different values of magnetic shear $\hat{s}$ (rows) in PT (left) and NT (right). The colorbar shows the value of the real frequency of the mode (where blue indicates MTM and yellow ITG), while the red dashed lines denote the nominal parameters of $R/L_{Ti}$ and $R/L_{Te}$ and the white squares are non-converged simulations.}
    \label{SMART_scanomte_omti}
\end{figure*}


\begin{figure*}[]
    \centering
    \includegraphics[width=\linewidth]{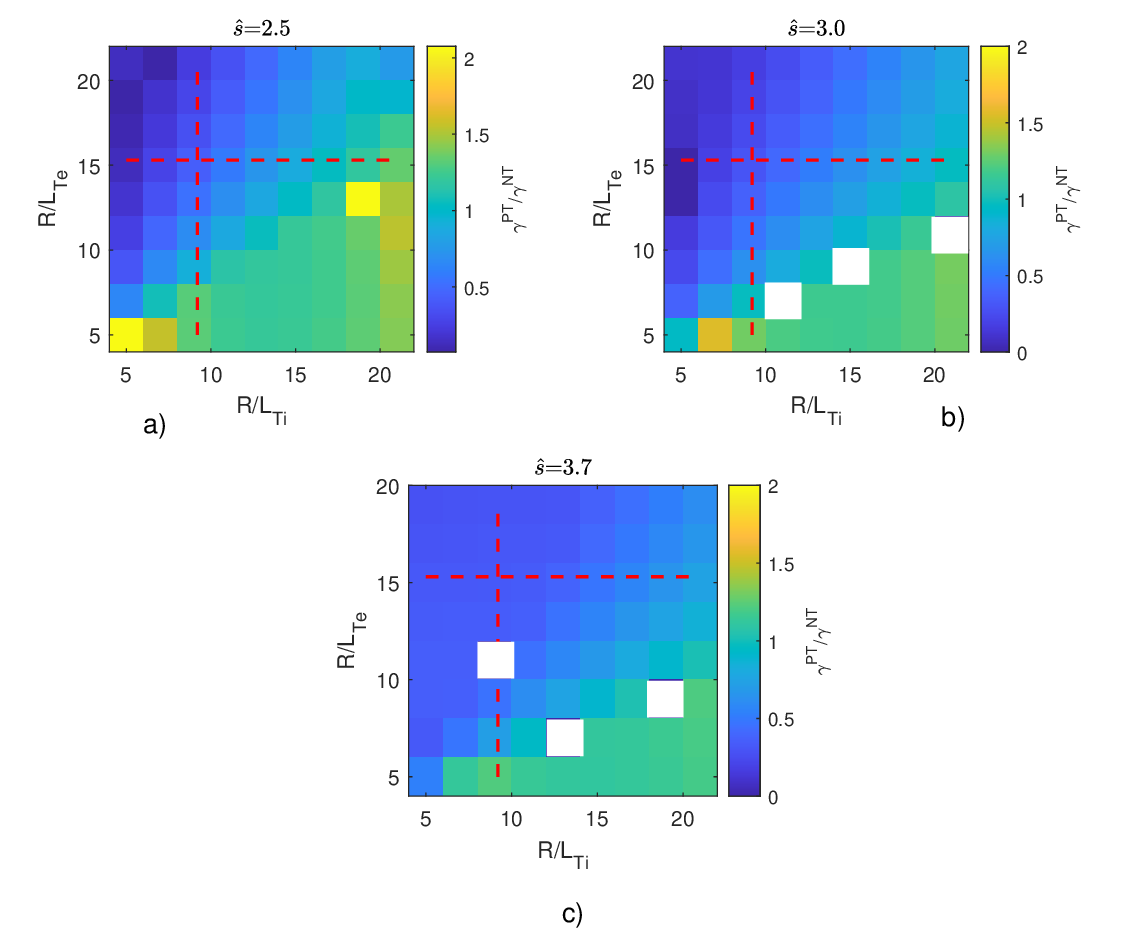}
    \caption{Two-dimensional scans of ion and electron temperature gradients for different values of magnetic shear $\hat{s}$. The colorbar shows the ratio of the growth rate in PT over the growth rate in NT geometry. A ratio larger than one means that NT is more stable than PT. The red dashed lines show the nominal parameters of $R/L_{Ti}$ and $R/L_{Te}$ and the white squares correspond to non-converged simulations.}
    \label{SMART_gamma_shear}
\end{figure*}

Next, we carried out a scan in collisionality for the $k_y\rho_i=0.4$ mode, which is shown in figure \ref{SMART_coll}. Again in this parameter scan we see a mode transition. At exactly zero collisionality, as predicted by previous theoretical work \cite{MTM}, we see that MTM cannot exist and instead the most unstable mode is TEM. For this TEM instability, NT has a lower growth rate than PT. However, as soon as the collisionality is finite, MTM dominates and NT has a faster growth rate. If we then increase collisionality, it reduces the gap between the growth rates of the two geometries. In figure \ref{SMART_omte}, we show a scan of the electron temperature gradient, which is another important drive of the MTM instability. Once again, we can see that PT and NT transition to MTM at somewhat different values, and that once the MTMs become dominant, the modes are more unstable in NT.

Another parameter that influences MTMs is the magnetic shear $\hat{s}$. Figure \ref{SMART_shear} shows the growth rate and frequency of the $k_y\rho_i=0.4$ mode as the magnetic shear is changed. We see that NT and PT have very different thresholds for the onset of the MTM regime. Indeed, NT transitions at a lower value of $\hat{s}$ for reasons not currently understood.

This linear GK analysis allowed us to identify the dominant type of turbulence in the PT and NT equilibria expected for SMART. In these equilibria both shapes should be dominated by MTM at the ion scale, and NT should have a stronger linear drive. However, we cannot make strong statements about the real performance of the NT option in SMART. Our linear simulations do not predict the nonlinear saturation of these modes and they did not consider the effect of impurities. However, based on the results obtained in this work, we believe that NT will be most beneficial for SMART in a regime where MTMs are less important and ITGs dominate. For this reason, we performed a triple scan in electron and ion temperature gradients and magnetic shear. The results are shown in figures \ref{SMART_scanomte_omti} and \ref{SMART_gamma_shear}.

First, we observe that an ITG region exists in both PT and NT and increases as the magnetic shear is reduced. In this region, where ITG turbulence dominates, linear growth rates are always lower in NT (the green regions in figure \ref{SMART_gamma_shear}). The opposite is true when we have MTMs. Therefore, it should be possible to operate in ITG-dominated conditions, where NT should be beneficial, but the size of this ITG region is considerably larger in PT than NT, making the requirements more strict. To conclude, if we want to operate SMART in a regime where we are confident to have a beneficial effect from NT, we should reduce the magnetic shear, increase the ion temperature gradient and decrease the electron temperature gradient.

\section{Physical picture of ITG at negative triangularity}\label{4}

As shown in sections \ref{3} and \ref{6}, the aspect ratio dependence of transport in NT and PT plasmas is complicated. To better understand it, we started by studying the scenario that we expected to be the simplest: the large $A$ pITG case. Indeed, in the limit of $A\gg$1 ITG modes depend only on two main geometry-related quantities: Finite Larmor Radius (FLR) effects and magnetic drifts. We then applied this picture to the conventional and tight $A$ cases, where many other factors arising from geometry can affect turbulence.

To make the derivation as clear as possible, we give here an outline of the contents of this section. First, we will illustrate how magnetic drifts and FLR effects influence the strength of ITG modes. Then, we will show how the geometry of the cross-sectional shape enters the FLR effects and magnetic drifts. This will allow us to show how NT weakens ITG. Lastly, we will show the numerical results that support this simple theory.

\subsection{Impact of magnetic drifts and FLR effects on ITG}\label{subA}

In this subsection, we will employ \cite{Beer} and \cite{Parisi_2020} to show how magnetic drifts and FLR effects contribute to the strength of ITG turbulence. 

Starting from the standard physical picture of the toroidal ITG instability \cite{Beer}, one can derive the relation 
\begin{equation}
    \omega_D=\frac{(\tau+b)\eta}{1+3(\tau+b)^2}\omega_*,
    \label{omegaD_res}
\end{equation}
that determines the condition for which the growth rate $\gamma$ of an ITG mode will be maximized. Here $\omega_D$ is the magnetic drift, $\tau=T_i/T_e$ is the ion to electron temperature ratio, $b=k_\perp^2\rho_i^2$,  $\rho_i$ is the ion Larmor radius, $k_\perp$ is the perpendicular wavenumber, $\eta=L_{ni}/L_{Ti}$, where $L_{ni}=-d(\ln{n_i})/dx$ is the ion density logarithmic gradient, $L_{Ti}=-d(\ln{T_i})/dx$ is the ion temperature logarithmic gradient, $\omega_*$ is the diamagnetic drift frequency. All the details regarding this derivation, including the definition of all the mentioned quantities, are reported in appendix \ref{ITG_appendix}.

If we set $\tau=1$ and $b=0$ and we recast the formula in terms of velocities, the condition becomes
\begin{equation}
    \textbf{v}_D=\frac{\eta}{4}\textbf{v}_*.
    \label{resonance_OK}
\end{equation}
This equation shows that to have the maximum possible growth rate, the magnetic drift and the diamagnetic drift velocities have to be of the same order. Since this is a very simplified model, one should not focus on the actual value of the constant that relates the two velocities. Rather, we should focus on the fact that if the magnetic drift velocity of the ions $\textbf{v}_D$ is too high or too low with respect to the diamagnetic velocity $\textbf{v}_*$, the growth rate will be small. In simple terms, if the depletion/accumulation zones of ions are traveling too fast or too slow with respect to the ion drift wave, a mismatch between the two develops, making the mode less unstable. The need to satisfy this ''resonance" is key in the explanation of NT beneficial effect on ITGs.

Apart from magnetic drifts, the only other way that the flux surface shape enters into the GK model in the $A\gg1$ limit is through the FLR effects. Specifically, in the local GK equations, the radial dependence of equilibrium quantities is dropped by the application of a gyroaverage, i.e. an integral average along the fast gyromotion of particles. It can be shown \cite{Merz} that the gyroaverage of a scalar field $\mathcal{F}$ reduces to a multiplication of the field with the zeroth order Bessel function $J_0$, i.e.: 
\begin{equation}
    <\mathcal{F}(\textbf{X})>=\sum_{k_y,k_x}\mathcal{F}(k_x,k_y)e^{i\textbf{k}_\perp\cdot\textbf{X}}J_0(\rho_\sigma k_\perp),
\end{equation}
where $k_x$ and $k_y$ are the radial and binormal Fourier modes and $\rho_\sigma$ is the Larmor radius of the $\sigma$ species. Therefore, the FLR effects correspond to a damping of the mode. The larger $\textbf{k}_\perp$, the stronger the damping.

\subsection{Impact of plasma shape on magnetic drifts and FLR effects}\label{subB}

FLR effects enter the GK model through the zeroth order Bessel function $J_0(k_\perp^2\rho_s^2)$. Since we are interested in the geometric dependence of this quantity, it is important to note that $\textbf{k}_\perp$ is the perpendicular wavenumber of a mode and is defined as
\begin{equation}
\textbf{k}_\perp=k_x\boldsymbol{\nabla}x+k_y\boldsymbol{\nabla}y.
\end{equation}
Therefore, using the quantities defined in appendix \ref{B}, we can write $k_\perp^2$ as
\begin{equation}
    k_\perp^2=g^{xx}k_x^2+2g^{xy}k_xk_y+g^{yy}k_y.
    \label{kperp2_OK}
\end{equation}
Hence, the geometry of the cross-sectional shape enters equation \eqref{kperp2_OK} through $g^{xx}$, $g^{xy}$ and $g^{yy}$. On the other hand, the magnetic drift frequency is $\omega_D=\textbf{v}_D\cdot\textbf{k}_\perp$. Using equation \eqref{vD}, we see it can be written as
\begin{equation}
\begin{split}
    \omega_D&=\left[\frac{1}{B\Omega}\left(\frac{v_\perp^2}{2}+v_\parallel^2\right)\textbf{b}\wedge\boldsymbol{\nabla}B\right]\cdot\left[ k_x\boldsymbol{\nabla}x+k_y\boldsymbol{\nabla}y\right]\\&=\Lambda\left(k_x\underbrace{\left(-\partial_y B-\frac{\gamma_2}{\gamma_1}\partial_z B\right)}_{v_{Dx}}+k_y\underbrace{\left(\partial_x B -\frac{\gamma_3}{\gamma_1}\partial_z B\right)}_{v_{Dy}}\right),
    \end{split}
    \label{omegaD_OK}
\end{equation}
where the scalar $\Lambda$ comprises all constant terms, $v_{Dx}$ is the radial component of the magnetic drift velocity and $v_{Dy}$ the binormal one. The derivation of this equation is reported in appendix \ref{C}. Therefore, the geometry of the plasma shape determines the drift frequency through the derivatives of the magnetic field and the elements of the metric tensor. Note that, with the GENE sign convention, a negative value of $\omega_D$ corresponds to the bad curvature region and positive values to the good curvature region. 

\begin{figure}[]
    \centering
    \includegraphics[width=0.99\linewidth]{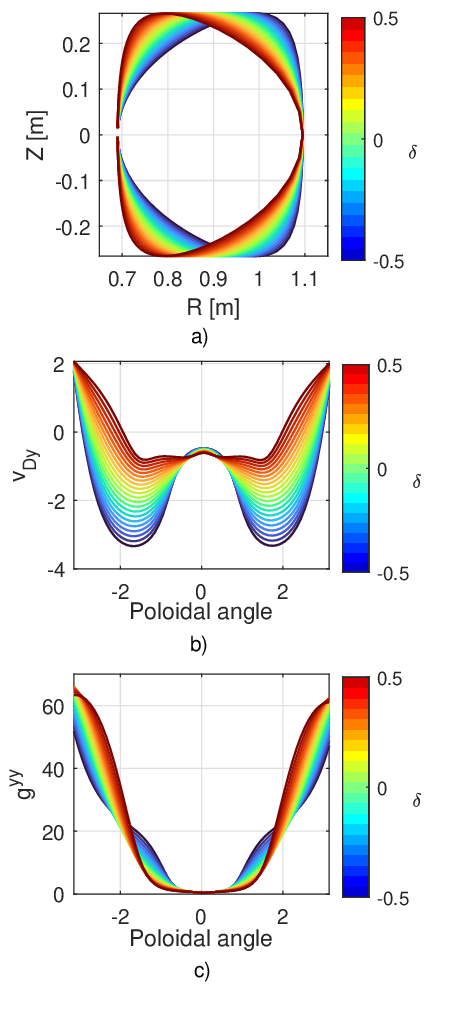}
    \caption{The (a) flux surface shape, (b) binormal component of the drift velocity $v_{Dy}$ as a function of the poloidal angle and (c) $g^{yy}$ as a function of the poloidal angle as the triangularity is varied.}
    \label{ITGgeom}
\end{figure}

\begin{figure*}[t]
    \centering
    \includegraphics[width=\linewidth]{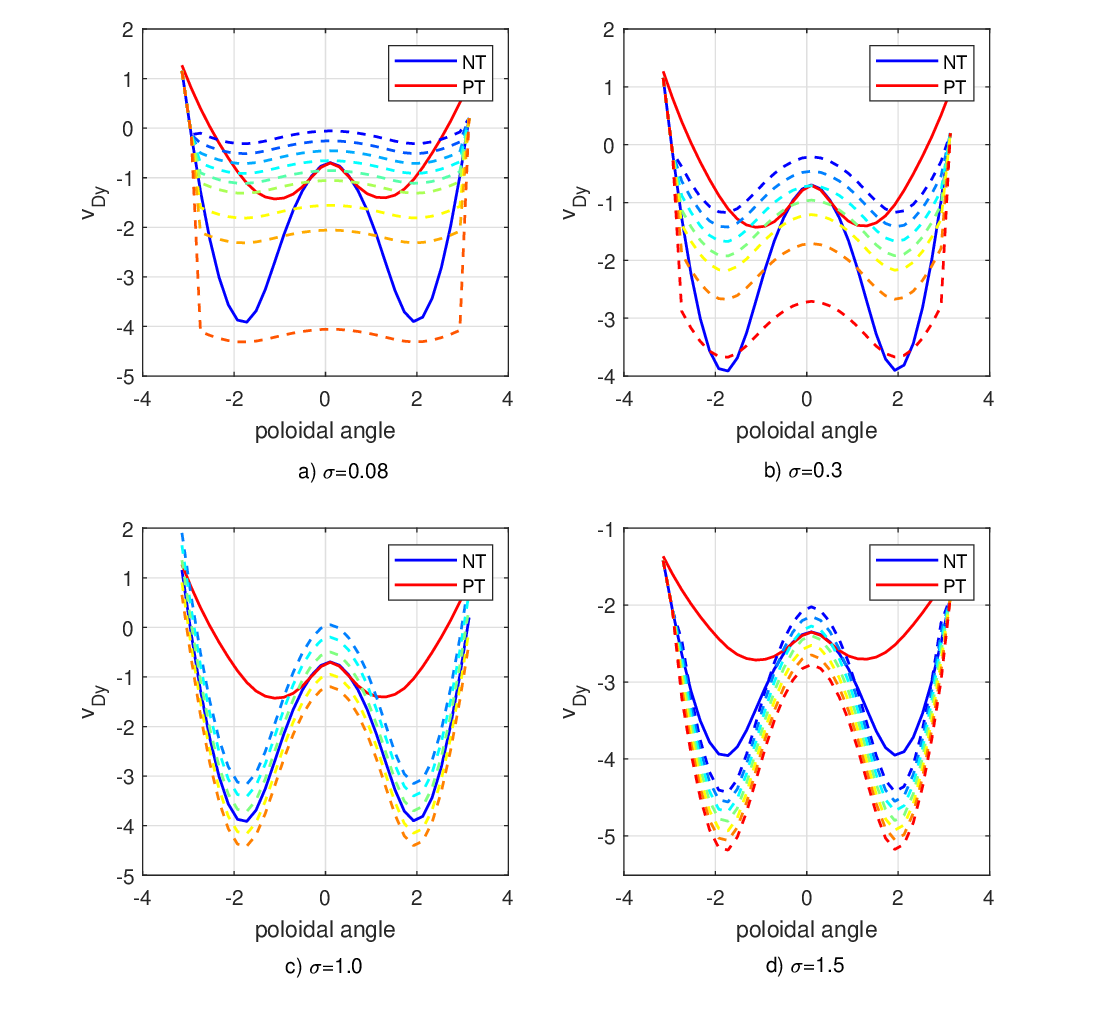}
    \caption{The ion magnetic drift velocity in NT (solid blue), PT (solid red), and NT profiles modified by $\sigma$ and $v_{Dy0}$ according to equation \eqref{vDy_mod}}
    \label{vDy_profs}
\end{figure*}

Now that we have expressions for the FLR effects and magnetic drifts, we can see how triangularity affects them. It must be pointed out that these two quantities, contrary to the simplified model presented in the previous subsection, depend on the poloidal angle, and are thus a function of $z$. Moreover, both FLR effects and magnetic drifts depend on $k_x$ and $k_y$. To simplify our analysis, we will consider the $k_x=0$ mode as it is typically the fastest growing linear mode for standard turbulence. Doing so allows us to test $k_y$ as a prefactor and will use only $g^{yy}$ as a proxy for the FLR effects and $v_{Dy}$ as a proxy for the magnetic drifts.

By using Miller local equilibrium geometry specification \cite{miller}, it is straightforward to calculate $g^{yy}$ and $v_{Dy}$ for any value of triangularity. In figure \ref{ITGgeom}(a) we take the equilibrium of the pITG-2 scenario and progressively change the triangularity. Figure \ref{ITGgeom}(a) shows the flux surface shapes, figures \ref{ITGgeom}(b) and \ref{ITGgeom}(c) show $v_{Dy}$ and $g^{yy}$ as functions of the poloidal angle for different values of triangularity. 

We see that, as the triangularity is decreased, the FLR effects become stronger in the vicinity of the outboard midplane, while the magnetic drifts become faster, i.e. more negative. In light of the observations made in the previous section, it is straightforward to say that one mechanism contributing to the stabilization of ITG in NT geometry comes from stronger FLR effects at the outboard midplane, where ITGs are stronger. However, the impact of magnetic drift is not clear. As mentioned before, negative values of $v_{Dy}$ correspond to bad curvature. However, more negative values do not necessarily mean stronger instability. Indeed, as we showed in the previous section, ITG is destabilized by matching the magnetic drift velocity and the diamagnetic drift velocity. Therefore, more negative values of $v_{Dy}$ can imply weaker ITG if $v_{Dy}$ is already larger than the condition set by equation \eqref{resonance_OK}. In the next section we will test this hypothesis with linear and nonlinear simulations.

\subsection{Numerical results}

We performed linear and nonlinear simulations to test if the faster magnetic drifts and stronger FLR effects of the NT geometry are responsible for weaker ITG turbulence.

\subsubsection{Linear results - Large $A$}

We started with linear GK simulations. The main goal was to verify that the simple  gyrofluid model presented in subsections \ref{subA} and \ref{subB} holds. 

We considered the large $A$ pITG-2 scenario and modified GENE to artificially change the poloidal dependence of $v_{Dy}$, keeping everything else untouched. We modified only $v_{Dy}$ because in linear simulations we consider only $k_x=0$ modes as they are assumed to grow the fastest. We took the original NT $v_{Dy}$ profile and modified it according to
\begin{equation}
    v_{Dy}(z)=\sigma v_{Dy}^{NT}(z)+v_{Dy0},
    \label{vDy_mod}
\end{equation}
where $\sigma$ and $v_{Dy0}$ are arbitrary scalar quantities that allowed us to change the variation of the profiles with poloidal angle and the offset respectively. We considered four values of $\sigma$ (i.e. 0.08, 0.3, 1.0 and 1.5) and for each of them we changed the offset to shift the profile up and down. The considered $v_{Dy}$ profiles are shown in figure \ref{vDy_profs}. For each modified profile, we performed a linear simulation for the $(k_x,k_y)=(0,0.4)$ mode, which is the fastest growing one and usually one of the modes that contributes the most to the NL heat fluxes. 

\begin{figure}
    \centering
    \includegraphics[width=\linewidth]{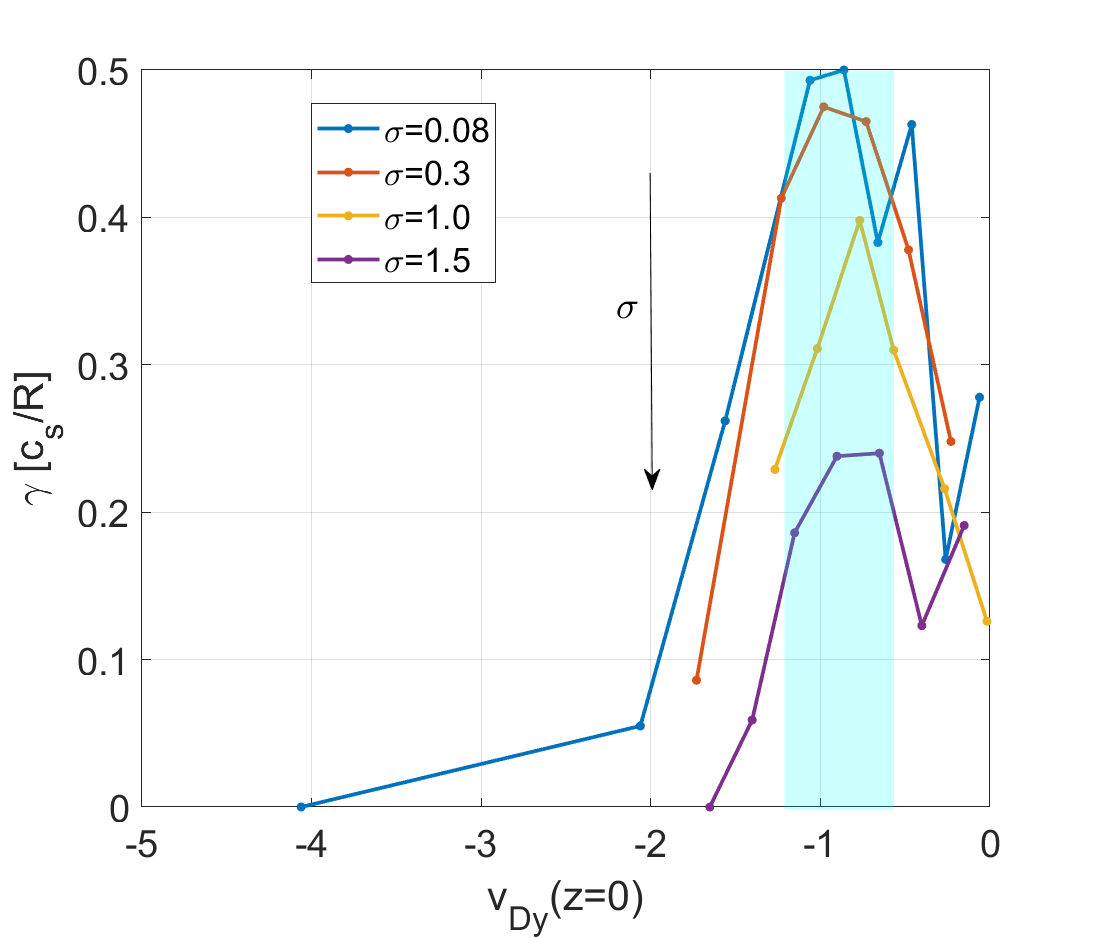}
    \caption{The growth rate $\gamma$ of the $k_x\rho_i=0$ $k_y=0.4$ mode for the different $v_{Dy}(z)$ functions shown in figure \ref{vDy_profs}}
    \label{gamma_vDy}
\end{figure}

This procedure allows us to isolate the effect of the magnetic drift and disentangle the contribution of the modulation from the value at the outboard midplane (where turbulence is strongest). Figure \ref{gamma_vDy} shows the results of the simulations. The different curves correspond to different modulations $\sigma$, each of which is plotted against the value of $v_{Dy}$ at $z=0$ in order to take into account the shift of the profile. We can see that the linear simulations confirm and deepen the simple picture that we gave at the beginning of the section. First, regardless of the modulation of the drift profile, a maximum point is always present at $v_{Dy}\sim-1$. Given the input parameters of our simulations $v_*\sim-13$ (in normalized units), so this optimal point is located at $v_{Dy}\sim v_*/13$ (as highlighted by the light blue box in figure \ref{gamma_vDy}). On the other hand, we see that the variation of $v_{Dy}$ with poloidal angle tends to stabilize the mode.

\begin{figure}
    \centering
    \includegraphics[width=\linewidth]{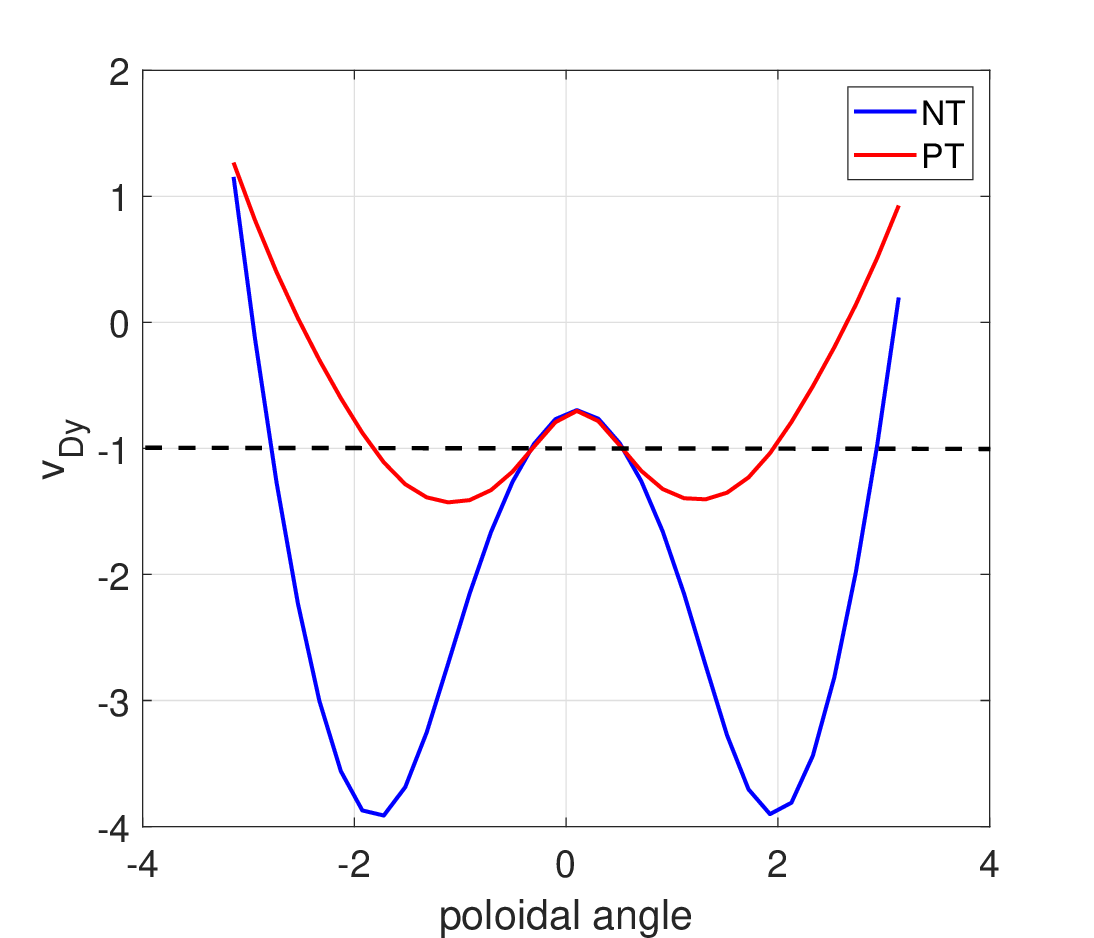}
    \caption{Binormal component of magnetic drift as a function of the poloidal angle for the NT and PT geometry. The dashed black line corresponds to the resonance condition $v_{Dy}=v_*/13$ found in the previous scan.}
    \label{resonance}
\end{figure}

\begin{figure*}[]
    \centering
    \begin{subfigure}[]
        { \includegraphics[width=\linewidth]{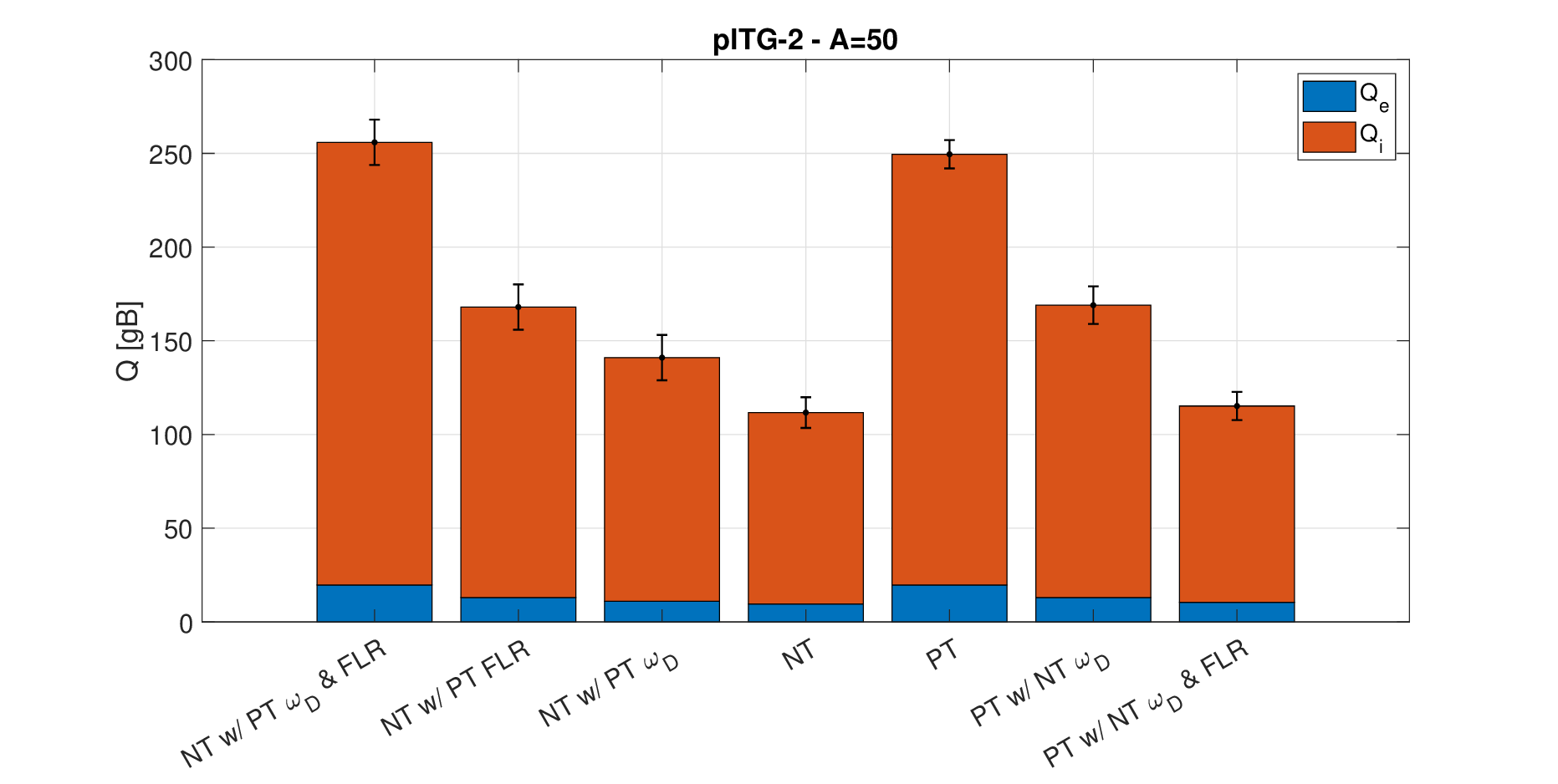}}
    \end{subfigure}
    \begin{subfigure}[]
        { \includegraphics[width=\linewidth]{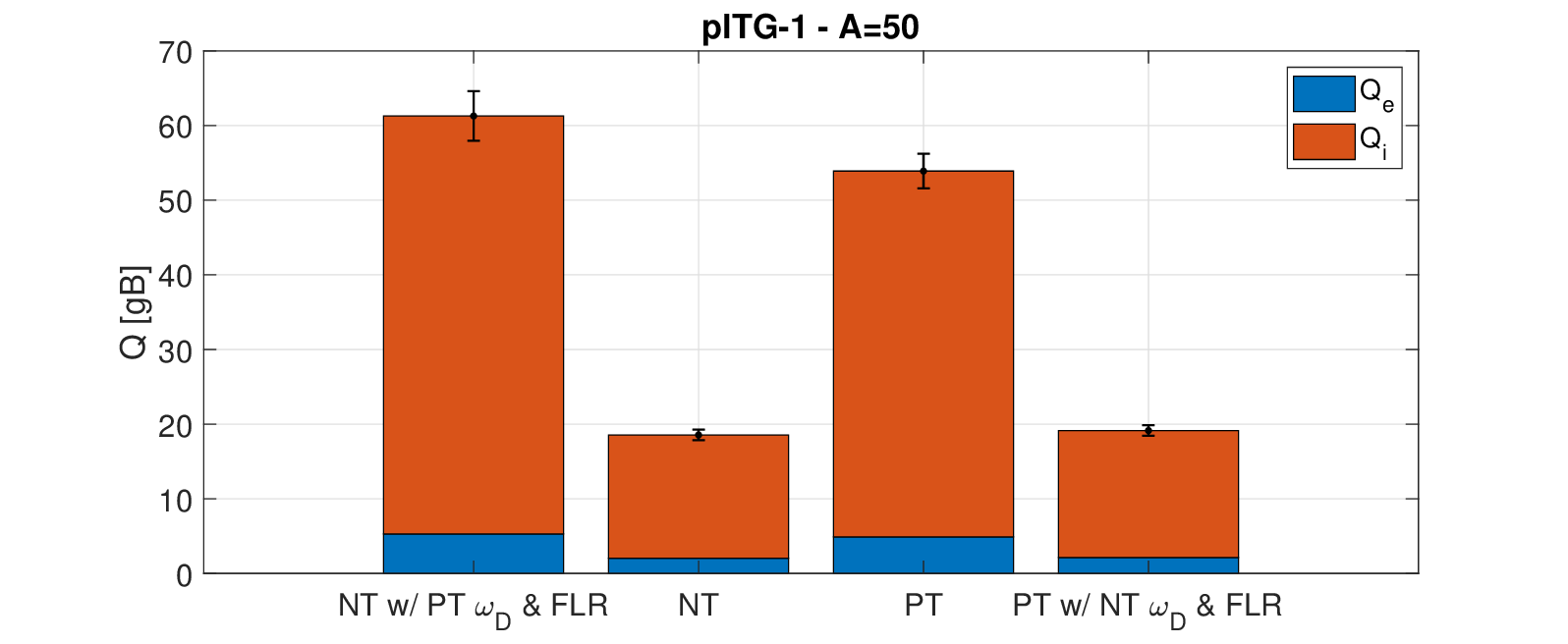}}
    \end{subfigure}
    \caption{The nonlinear heat fluxes in the (a) pITG-2 and (b) pITG-1 cases with a large aspect ratio of $A=50$.}
    \label{barQ_largeA}
\end{figure*}
With this information we can better understand how geometry influences ITG via magnetic drifts. In figure \ref{resonance}, we plot the ideal $v_{Dy}(z)$ to enable the resonance together with the actual $v_{Dy}(z)$ in the NT and PT cases. We see that $v_{Dy}$ varies less with poloidal angle in a PT geometry than in a NT one, and it is always closer to the resonance. Thus, the proximity to the resonance destabilizes ITG modes in PT geometry, explaining why NT is beneficial. However, linear simulations cannot be taken as a final proof. We also need nonlinear simulations to test our understanding in a realistic quasi-stationary turbulent state.

\subsubsection{Nonlinear results - Large $A$}

In this subsection, we show a series of nonlinear simulations to confirm that, at large A, stronger FLR effects and faster magnetic drifts fully explain the beneficial effect of NT on pure-ITG turbulence.

Nonlinear GK simulations are much more computationally expensive than linear simulations, which prevents an analogous scan to figure \ref{gamma_vDy}. Instead, we artificially swap all the geometric coefficients responsible for FLR effects and magnetic drifts between NT and PT. Indeed, from equations \eqref{kperp2_OK} and \eqref{omegaD_OK}, we can easily identify all the coefficients that one needs to exchange between the two geometries. Clearly, this procedure creates unrealistic equilibria, but it is perfect to isolate the contributions coming from FLR effects and magnetic drifts. Apart from these two effects, everything else is left untouched in the GK equations. A similar exercise was carried out by Merlo and Jenko in \cite{merlo_jenko_2023}, but only for linear cases.

We will consider the large $A$ pITG-2 scenario and perform new simulations where FLR effects or magnetic drifts $v_D$ (or both) are changed. All simulations were performed with a kinetic treatment of electrons. Figure \ref{barQ_largeA}(a) displays the results. The fourth and fifth color bars of figure \ref{barQ_largeA}(a) correspond to the self-consistent reference NT and PT cases. From the third color bar, we can see that, by imposing PT's $v_{D}$ in the NT geometry, we can increase the heat flux by a factor of 1.3 . From the second column, we can see that by changing only the FLR effects we can increase the heat flux by a factor of 1.5. This is a very important result, because it confirms our simple physical picture showing that that FLR and magnetic drifts are destabilizing in PT and they have similar importance. If we then look at the first column, i.e. the one where both coefficients were changed, we see that the new heat fluxes closely match the ones of the self-consistent PT case. As expected, this confirms that in the large aspect ratio limit turbulence can be described just in terms of FLR effects and magnetic drifts. The same arguments are supported by columns six and seven. We also performed the same exercise for the large $A$ pITG-1 scenario, which is still characterized by pITG turbulence, but has different input parameters and magnetic equilibria. The results are shown in figure \ref{barQ_largeA}(b) and they show the very same findings observed before. These observations confirm that, in the large $A$ limit, the beneficial effect of NT on ITG comes from a stabilizing effect arising from stronger FLR effects and faster magnetic drifts with respect to the resonance condition set by the diamagnetic drift velocity and thus by the ion temperature gradient. 

Encouraged by these results, we tried to apply our physical picture to the conventional and tight $A$ pITG cases and to other large $A$ pITG cases where other geometric parameters were changed apart from triangularity. In the following subsection, we show the results for the conventional and tight $A$ pITG cases, where we applied the very same methodology showed before. In appendix \ref{D} we show the results for scans in other geometric parameters (i.e. elongation and magnetic shear).

\subsubsection{Nonlinear results - Conventional and small A}

\begin{figure}[]
    \centering
    \includegraphics[width=0.98\linewidth]{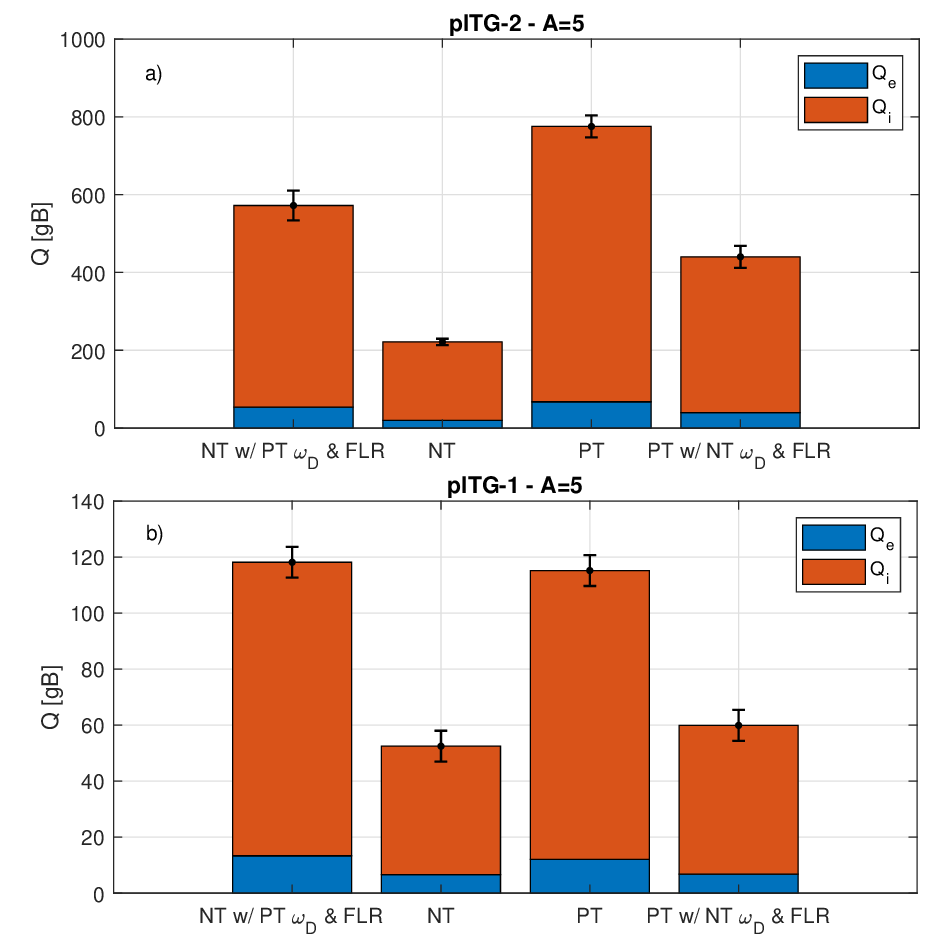}
    \caption{The nonlinear heat fluxes in the (a) pITG-2 and (b) pITG-1 cases with a conventional aspect ratio $A=5$.}
    \label{barQ_A}
\end{figure}

We performed the same numerical study from the previous section, but with conventional A. The results of nonlinear simulations for pITG-2 and pITG-1 cases are shown in figure \ref{barQ_A}. We see that stronger FLR effects and faster magnetic drifts are also stabilizing at this aspect ratio. However, at finite aspect ratio the geometry enters via many more effects (e.g. particle trapping, parallel streaming), so it is not necessarily true that these two effects alone can explain the beneficial effect of NT. We see that for both scenarios, FLR effects and magnetic drifts have a dominant effect on the heat fluxes. For the pITG-1 case, they can entirely explain the stabilizing effect of NT. For the pITG-2 case, they account for approximately two-thirds of the stabilization. Therefore, also in the conventional aspect ratio cases, stronger FLR effects and faster magnetic drifts explain the beneficial effect of NT.


\begin{figure}[]
    \centering
    \includegraphics[width=0.98\linewidth]{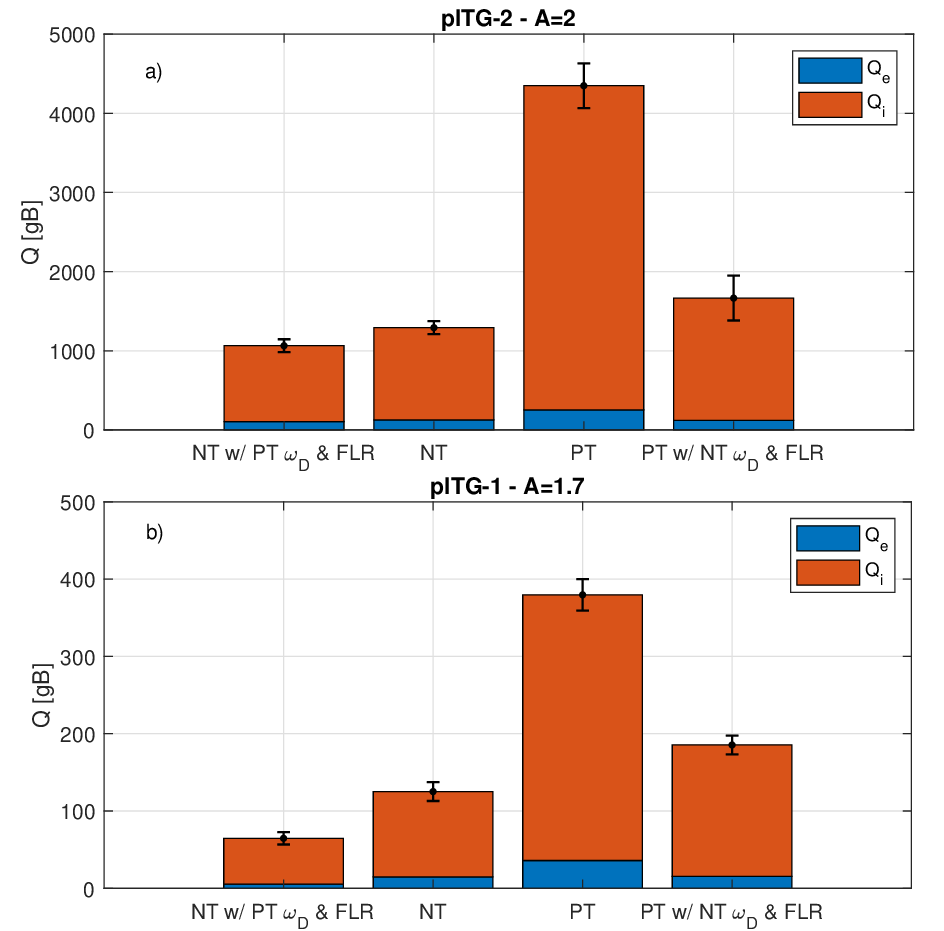}
    \caption{The nonlinear heat fluxes in the (a) pITG-2 and (b) pITG-1 cases with tight aspect ratios $A=2.0$ and $A=1.7$ respectively.}
    \label{barQ_lowA}
\end{figure}

\begin{figure*}[]
    \centering
    \includegraphics[width=\linewidth]{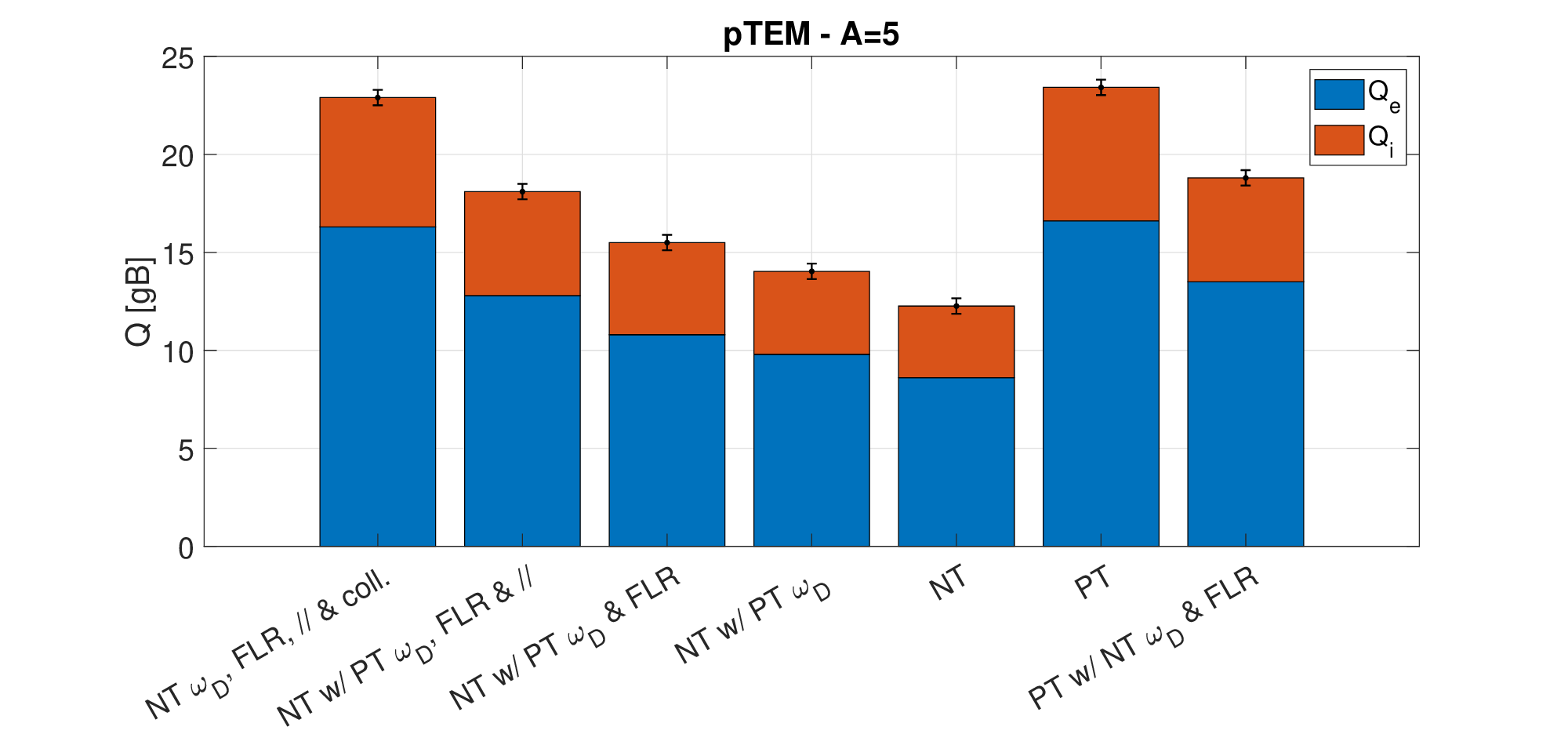}
    \caption{The nonlinear heat fluxes in the pTEM case with a conventional aspect ratio of $A=5$.}
    \label{barQ_pTEM}
\end{figure*}

As a final remark, we also investigated to what extent our physical picture can explain the effect of NT at tight A. We can see from figure \ref{barQ_lowA} that magnetic drifts and FLR effects are no longer sufficient to explain the stabilization of ITG turbulence by NT. Contrary to what was shown before, we see complicated and counter-intuitive results. When we impose the FLR effects and magnetic drifts from NT into the PT geometry, turbulence is reduced as in previous cases with $A=50$ and $A=5$. On the other hand, when we impose the FLR effects and magnetic drifts from PT into the NT geometry, we still see a stabilization. These trends are clearly in contrast with the physical picture that we held for large and conventional A, suggesting that at tight aspect ratio other physical processes are involved. The investigation of these is left as future work.

\section{Physical picture of TEM at negative triangularity}\label{5}

TEMs are driven unstable by a resonance in phase space between the diamagnetic drift velocity of the underlying drift wave and the toroidal precessional drift of trapped electrons. Based on this fundamental drive, one can construct a similar physical picture as we did for ITGs. This has already been done in \cite{Marinoni_2009}. It can be shown that the toroidal precessional drift is closely related to the bounce average of the magnetic drift velocity of trapped particles \cite{Graves_2013}.

In this section, we will evaluate the NT TEM physical picture considered in the aspect ratio scan. We had to consider a conventional aspect ratio scenario to have sufficiently strong TEM instability. At large aspect ratio, the fraction of trapped particles is very low, so the strength of TEM turbulence is very weak. This complicates the physical picture because, as already mentioned, only at large aspect ratio we can reduce the effect of geometry just to FLR effects and magnetic drifts. However, even if we cannot explain everything in terms of these two factors, it is important to understand the significance of their role in the stabilization of TEMs by NT.

We proceeded with a similar methodology as used for the NL simulations of the pITG cases: we swapped the magnetic drifts, FLR effects and other effects between NT and PT.

Figure \ref{barQ_pTEM} shows the results. The fifth and sixth color bars show the NT and PT self-consistent cases, i.e. all the physical factors are computed self-consistently with the respective geometry. The fourth column shows the case where only magnetic drifts have been swapped between NT and PT. We can see that the heat fluxes rise, but the increase is mild. If we move further to the left, the next bar corresponds to the case where the FLR effects have been swapped along with magnetic drifts. The heat fluxes are increased further, but the total heat flux is still much lower than in the PT case. The same trend can be observed if we do the same exercise starting from the PT scenario, as shown by the seventh bar. Therefore, we can conclude that, in TEM-dominated regimes, faster magnetic drifts and stronger FLR effects are only some of the factors contributing to the reduction of turbulence by NT. To find the missing pieces of the puzzle, we swapped other geometric factors that could play a role in the stabilization of TEM turbulence. The results correspond to the first and second bar of figure \ref{barQ_pTEM}. First, we swapped the geometric parameters responsible for the parallel dynamics, i.e. the parallel advection term and the trapping term \cite{3}. The parallel dynamics can play an important role in TEM turbulence because they are related to the bouncing points of the trapped particles. Swapping the parallel dynamics (in addition to the magnetic drifts and FLR effects) is showed by the second column. We observe that this increases the total heat flux, but it is still lower than in the PT case. Therefore, the parallel dynamics are part of the story, but not the most important factor for the stabilization of TEM by NT. We also tried to swap the geometric parameters that enter the computation of the collisional operator (when expressed in parallel velocity $v_\parallel$ and magnetic moment $\mu$ coordinates), namely the magnetic field strength and the Jacobian. This effect is important because it regulates the diffusion of particles across the trapping-passing boundary, influencing the fraction of trapped particles and thus TEM turbulence. Accordingly, if we look at the first column, where we have also swapped the collisional diffusion between NT and PT, we can see that it is possible to recover the same heat flux of the self-consistent PT case. Moreover, the change in the heat flux is larger than that caused by all the previous geometrical effects.

For the pTEM case, we can conclude that the physics underlying the confinement benefits of NT is not due to a single physical effect but instead a synergistic combination. This exercise is similar to previous work \cite{Marinoni_2009}, but the results are slightly different, potentially due to the different equilibria considered. Indeed, \cite{Marinoni_2009} was able to recover the heat flux of the opposite triangularity scenario just by swapping magnetic drifts and FLR effects. Here, they explain only part of the picture.

To conclude, in this section we showed that FLR effects and the resonance between magnetic drifts and diamagnetic drifts can be applied, to some extent, to explain how NT stabilizes TEM turbulence. However, they alone are not sufficient to explain the entire effect. Parallel dynamics and collisional diffusivity also play a crucial role.

\section{Conclusions}\label{7}

This work consisted of a thorough numerical investigation of the interplay between aspect ratio and the beneficial effect of NT on turbulent transport. It consisted of three distinct, but complementary studies.
\begin{itemize}
    \item In the first part, we considered five scenarios characterized by different types of turbulence: a realistic TCV-inspired TEM-dominated scenario, a realistic DIII-D-inspired ITG-dominated scenario, two idealized pure-ITG scenarios, and an idealized pure-TEM scenario. We observed that at large and conventional aspect ratio, NT had better confinement than PT regardless of the turbulent regime. At tight aspect ratios, for ITG-dominated turbulence, NT remained beneficial with respect to PT, while NT was detrimental in TEM-dominated scenarios, leading to heat fluxes larger by a factor of up to 2.5. An analysis of the stiffness showed that at conventional aspect ratio, flipping the sign of triangularity from PT to NT increases the critical gradient and leaves the stiffness unaffected. For small A, flipping the triangularity does not influence the critical gradient, but can change the stiffness. Specifically, for TEM NT is stiffer, while for ITG scenarios PT is stiffer.
    \item In the second part, we performed the first GK simulations of the NT and PT equilibria predicted for SMART. Linear analysis showed that MTMs dominate at ion scales. Moreover, in NT they have larger linear growth rate when compared to PT with the same kinetic parameters. A scan in electron $\beta$ showed that MTMs appear at lower $\beta$ and $R/L_{Te}$ in NT and have larger growth rates. However, when ITG is the dominant type of instability, NT has weaker linear growth rates, which is compatible with the findings of the previous section. Finally, a scan of the magnetic shear showed that large values are crucial to drive MTMs at large $\beta$ and electron temperature gradients. A triple scan in ion temperature gradient, electron temperature gradient and magnetic shear showed that by lowering the magnetic shear and the ratio of electron temperature gradient to ion temperature gradient it is possible to extend the space where SMART NT option can be operated in ITG-dominated regime and exploit the beneficial effect of NT.
    \item In the third part, we started from a simplified model of the toroidal ITG instability and were able to explain the beneficial effect of NT on confinement for ITG at large and conventional A. Using linear and nonlinear simulations, we showed that the stabilization of ITG in NT comes from \textit{faster magnetic drifts} with respect to a resonance condition set by the diamagnetic drift, and from \textit{stronger FLR effects}. However, even though the beneficial effect of NT is still present at small $A$ in ITG regimes, the physical picture developed in this paper does not apply as many other physical effects are important.
\end{itemize}
We conclude that the interplay between aspect ratio and NT is complicated, but must be taken into account for the successful optimization of a possible NT fusion reactor. Based on our observations, the strong beneficial effect of NT can be exploited at any aspect ratio if we operate in an ITG-dominated regime. However, based on the result for SMART, we conclude that in a ST, MTMs may well be the dominant instability. If so, NT will likely have more unstable turbulence than PT. However, NL simulations are needed to confirm this picture. Future work will focus on understanding if the observations made for MTM turbulence in spherical tokamaks hold in standard tokamaks as well.

\section{Data availability statement}

The data that support the findings of this study are available upon reasonable request from the authors.

\section*{Acknowledgments}

We acknowledge Jason Parisi for the extremely insightful conversation on curvature drive for ITG and ETG turbulence, which paved the way for the explanation of why NT stabilizes ITGs. We acknowledge Stephan Brunner for all the invaluable suggestions and for pointing us to M. Beer's Ph.D thesis. We also thank Giovanni Di Giannatale, Alessandro Geraldini, Antoine Hoffmann, Fabien Jeanquartier, Oleg Krutkin, Haomin Sun and Arnas Volcokas for the constant help and inspiring discussions. This work has been carried out within the framework of the EUROfusion Consortium, via the Euratom Research and Training Programme (Grant Agreement No 101052200 — EUROfusion) and funded by the Swiss State Secretariat for Education, Research and Innovation (SERI). Views and opinions expressed are however those of the author(s) only and do not necessarily reflect those of the European Union, the European Commission, or SERI. Neither the European Union nor the European Commission nor SERI can be held responsible for them. This work was also carried out within the Theory, Simulation, Verification and Vailidation (TSVV) task 2. The authors acknowledge the CINECA award under the ISCRA initiative and the EUROfusion WP-AC, for the availability of high-performance computing resources and support. This work was also carried out (partially) using supercomputer resources provided under the EU-JA Broader Approach collaboration in the Computational Simulation Centre of International Fusion Energy Research Centre (IFERC-CSC). This work was supported in part by the Swiss National Science Foundation. 

\appendix

\section{Linear simulations to establish turbulent regime}\label{A} 

\begin{figure*}[]
    \centering
    \begin{subfigure}[Growth rate]
        { \includegraphics[width=0.3\linewidth]{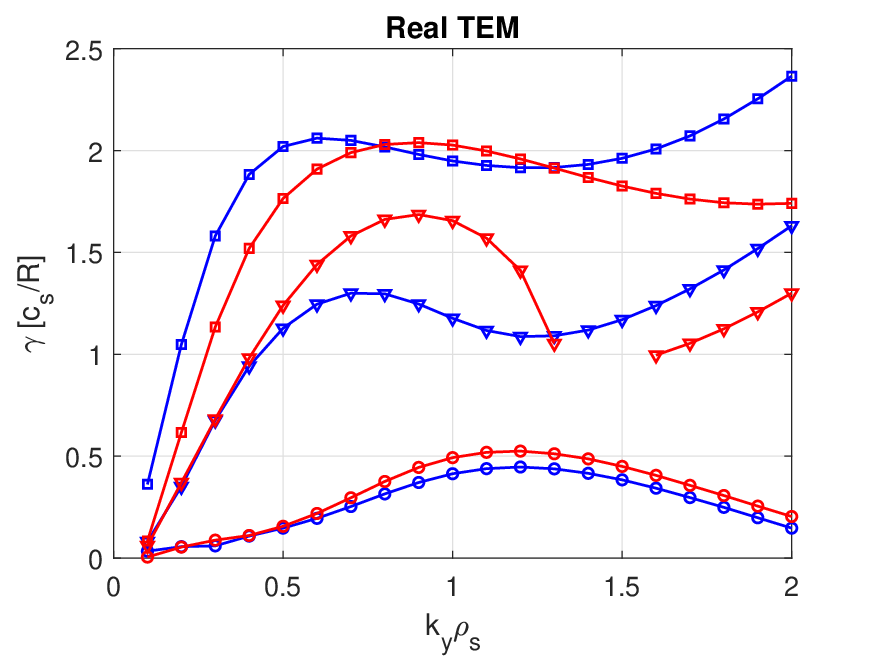}}
    \end{subfigure}
    \begin{subfigure}[Frequency]
        { \includegraphics[width=0.3\linewidth]{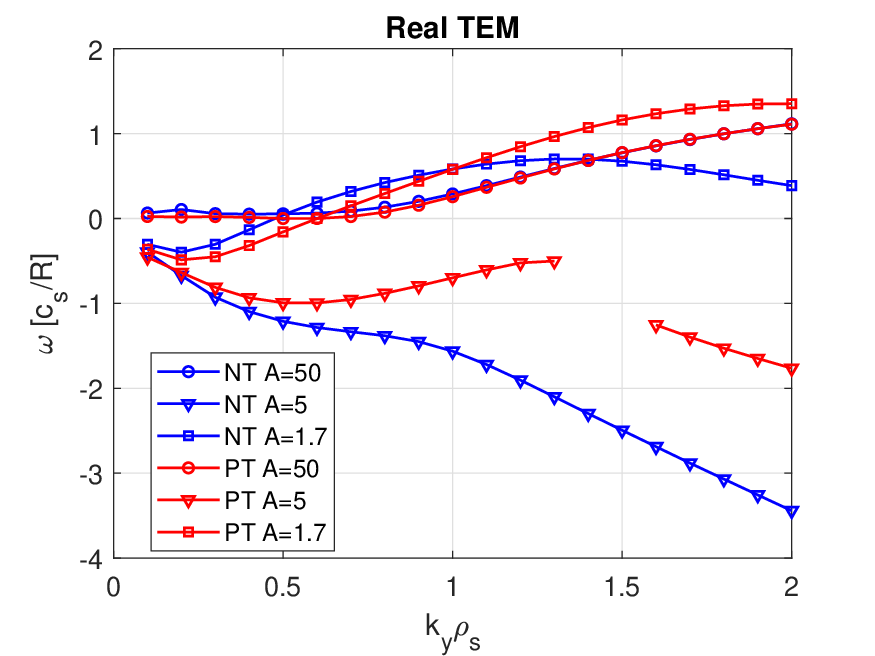}}
    \end{subfigure}\\
    \begin{subfigure}[Growth rate]
        { \includegraphics[width=0.3\linewidth]{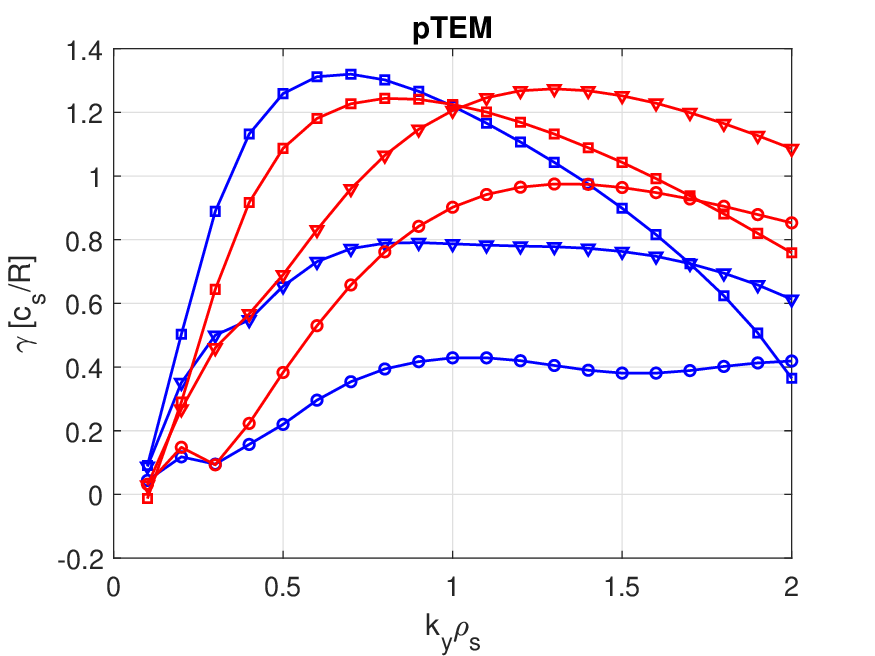}}
    \end{subfigure}
    \begin{subfigure}[Frequency]
        { \includegraphics[width=0.3\linewidth]{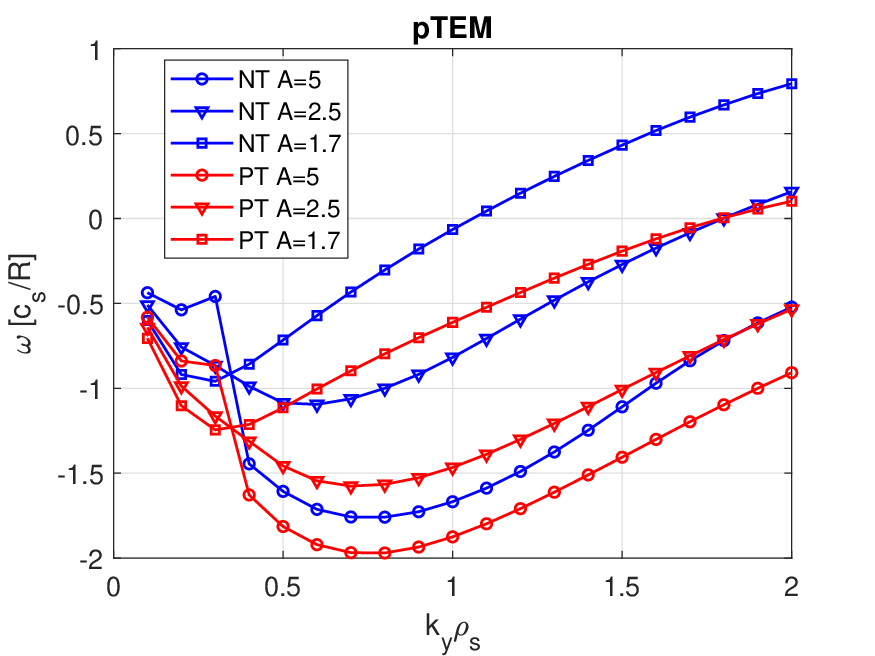}}
    \end{subfigure}\\
    \begin{subfigure}[Growth rate]
        { \includegraphics[width=0.3\linewidth]{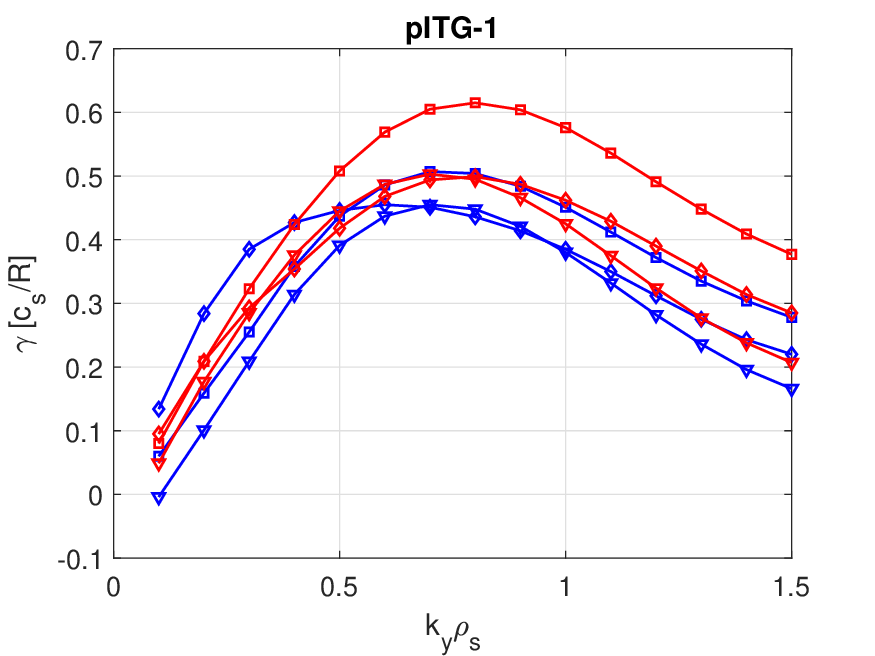}}
    \end{subfigure}
    \begin{subfigure}[Frequency]
        { \includegraphics[width=0.3\linewidth]{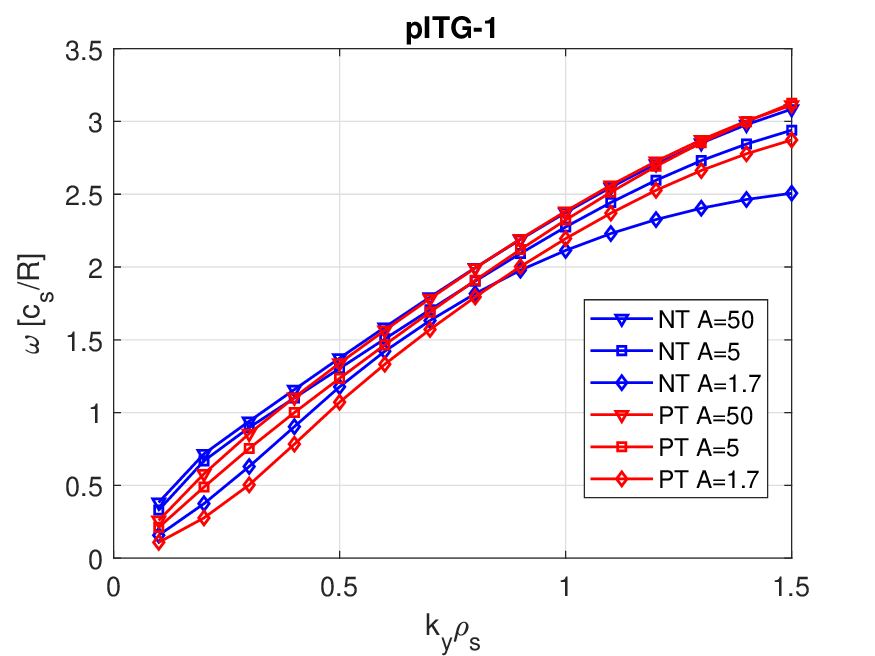}}
    \end{subfigure}\\
    \begin{subfigure}[Growth rate]
        { \includegraphics[width=0.3\linewidth]{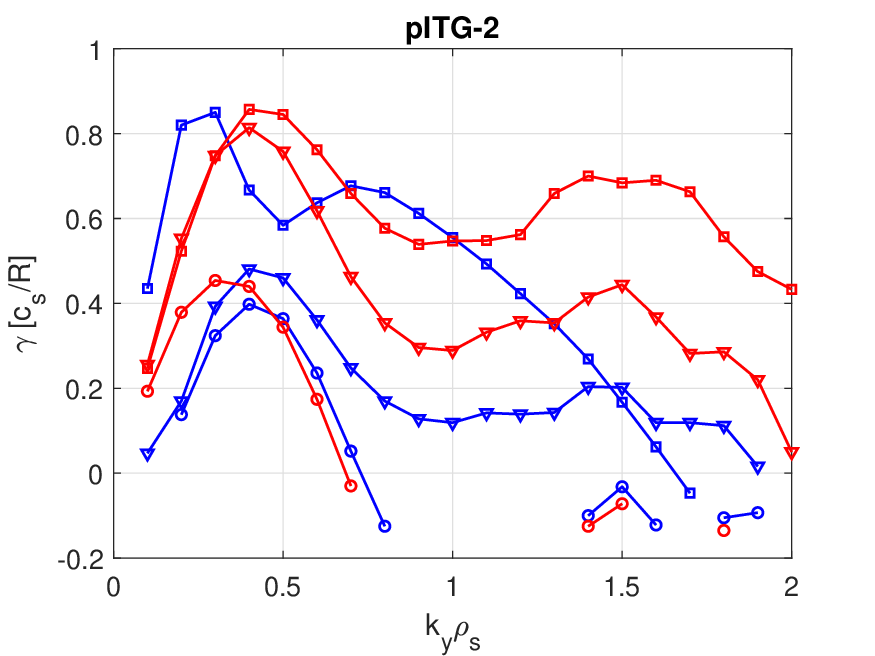}}
    \end{subfigure}
    \begin{subfigure}[Frequency]
        { \includegraphics[width=0.3\linewidth]{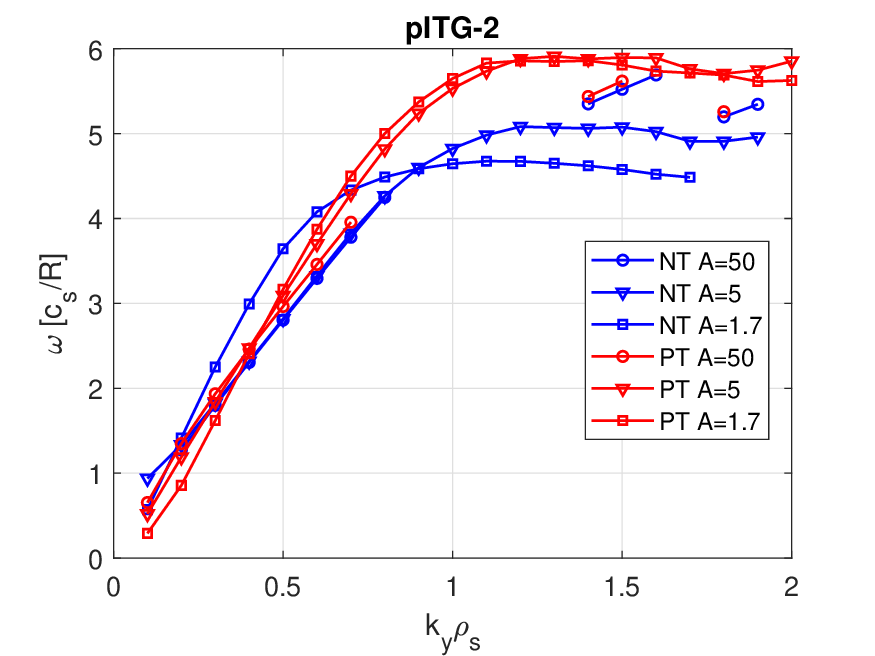}}
    \end{subfigure}\\
    \begin{subfigure}[Growth rate]
        { \includegraphics[width=0.3\linewidth]{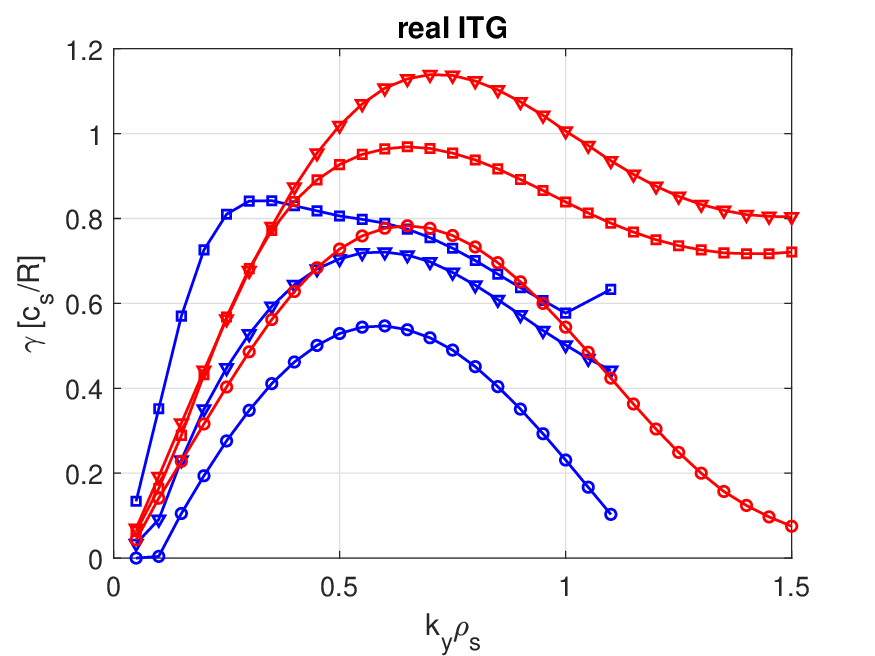}}
    \end{subfigure}
    \begin{subfigure}[Frequency]
        { \includegraphics[width=0.3\linewidth]{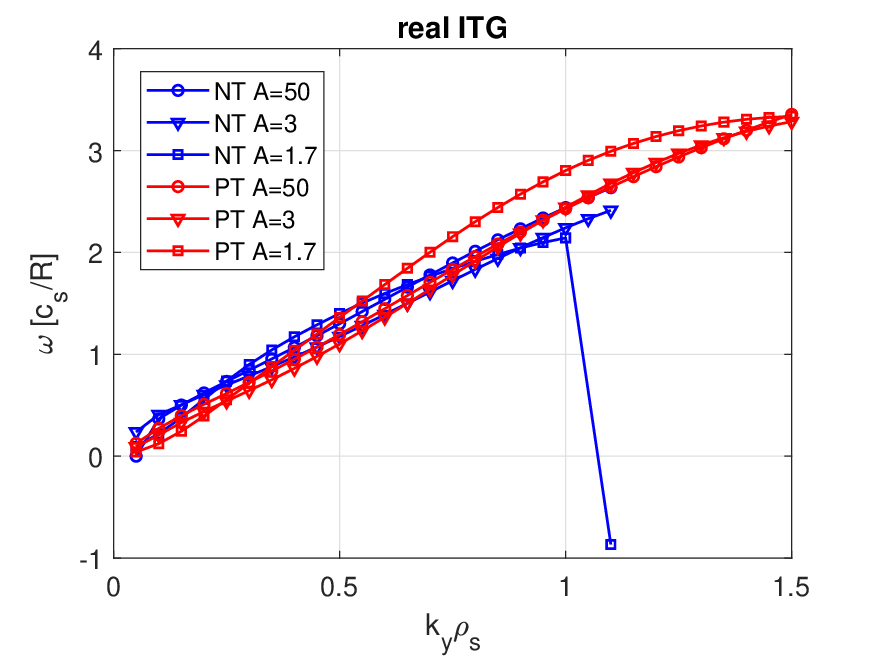}}
    \end{subfigure}
    \caption{ Growth rates (on the right) and real frequencies (on the left) as functions of the binormal wavenumber in different turbulent regimes and at different aspect ratios.}
    \label{linear_A}
\end{figure*}

In this appendix, we show the results of the linear simulations performed for the different scenarios shown in figure \ref{Qratio_A} to determine the turbulent regime as the aspect ratio is changed. Figure \ref{linear_A} shows the growth rates and the frequencies as functions of the binormal wavenumber for the $k_x=0$ mode. We see that, as expected, the real TEM and pTEM cases are dominated by TEM turbulence while the pITG-1, pITG-2 and real ITG are dominated by ITG turbulence.

\section{Derivation of the condition for the optimum growth rate of an ITG mode\label{ITG_appendix}
}
\begin{figure*}
    \includegraphics[width=\linewidth]{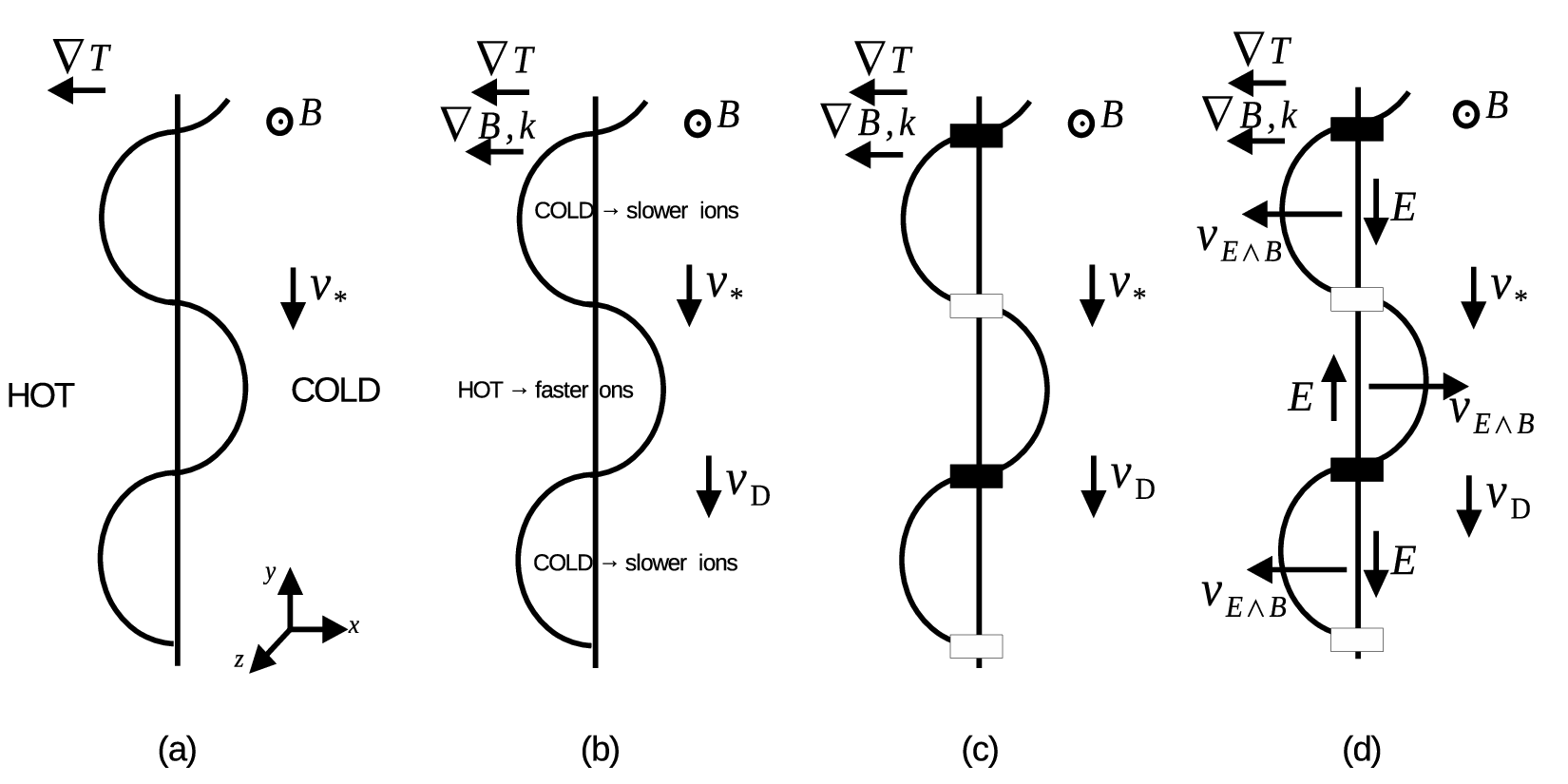}
    \caption{Qualitative picture of the formation of an unstable ITG mode.}
    \label{ITG}
\end{figure*}
In this appendix we report the derivation that leads to equation \eqref{omegaD_res}. Toroidal ITG modes are driven by the magnetic field line curvature and gradients in the magnetic field strength. They can be unstable where the curvature is ``bad'', meaning the magnetic field curvature (or equivalently gradient) has a component parallel to the temperature gradient.

Let's consider a simplified slab domain, where the magnetic field points along $z$ and the temperature gradient points in the $-x$ direction. This would correspond to the outboard midplane in a tokamak and is shown in figure \ref{ITG}(a). A small perturbation of the ion pressure will generate a neutrally stable ion drift wave that moves downwards with a velocity $\textbf{v}_*=\frac{\textbf{B}\wedge\boldsymbol{\nabla}P}{qNB^2}$, also called \textit{diamagnetic drift velocity}, where $P$ is the pressure, $q$ the charge of the ion, $N$ the density and $\textbf{B}$ is the magnetic field. Now, if we include the presence of the magnetic field curvature $\boldsymbol{\kappa}$ and gradient $\boldsymbol{\nabla}B$, the ions will drift with a velocity $\textbf{v}_D$, which is called the \textit{magnetic drift velocity}. It is the sum of two terms, one from the curvature $\textbf{v}_c$ and one from the gradient $\textbf{v}_{\nabla B}$, and is expressed as
\begin{equation}
    \begin{split}
    \textbf{v}_D&=\textbf{v}_{c}+\textbf{v}_{\nabla B}\\&=\frac{v_\parallel^2}{\Omega}(\boldsymbol{\nabla}\wedge \textbf{b})_\perp+\frac{v_\perp^2}{2B\Omega}\textbf{b}\wedge\boldsymbol{\nabla}B\\&=\frac{v_\parallel^2}{\Omega B}(\textbf{b}\wedge(\boldsymbol{\nabla}B+\frac{4\pi}{B}\boldsymbol{\nabla}P)+\frac{v_\perp^2}{2B\Omega}\textbf{b}\wedge\boldsymbol{\nabla}B,
    \end{split}
\end{equation}
where $\textbf{b}=\textbf{B}/B$ is the unit vector parallel to $\textbf{B}$, $\Omega$ is the Larmor frequency, and $v_\parallel$ and $v_\perp$ are the components of the velocity parallel and perpendicular to the magnetic field respectively. If we neglect pressure gradient (i.e. in the $\beta=P/B^2/2\mu_0\ll1$ limit), we can write for simplicity
\begin{equation}
    \begin{split}
    \textbf{v}_D&=\frac{1}{B\Omega}\left(\frac{v_\perp^2}{2}+v_\parallel^2\right)\textbf{b}\wedge\boldsymbol{\nabla}B.
    \end{split}
\end{equation}
We can estimate the characteristic value of this velocity as
\begin{equation}
    \textbf{v}_D=\frac{v_{th}^2}{B\Omega}\textbf{b}\wedge\boldsymbol{\nabla}B,
    \label{vD}
\end{equation}
where $v_{th}=\sqrt{T/m}$ is the ion thermal speed. Therefore, at the outboard midplane where $\boldsymbol{\kappa}$ and $\boldsymbol{\nabla}B$ point in the $-x$ direction the ions will start drifting downwards as pictured in figure \ref{ITG}(b).

The temperature dependence in the expression for $\textbf{v}_D$ will make the ions in the cold regions drift slower than those in the warmer regions, creating zones of accumulation and depletion of ions. This mechanism is showed in figure \ref{ITG}(c). The subsequent charge separation creates an electric field that leads to another type of drift, the ExB drift, given by
\begin{equation}
    v_{E\wedge B}=\frac{\textbf{E}\wedge\textbf{B}}{B^2}.
\end{equation}
The direction of the ExB drift is crucial to determining stability. Indeed, at the outboard midplane, the ExB drift will reinforce the initial perturbation making the mode unstable, as illustrated in figure \ref{ITG}(d). Instead, if we are on the inboard midplane, the direction of the drift will be such to counter the initial perturbation, stabilizing the mode. For this reason, these two regions are called regions of ``bad" and ``good" curvature respectively.

One can derive a dispersion relation for the ITG modes, which corresponds to the qualitative physical picture \cite{Beer}. In gyrofluid theory, it is given by. 
\begin{equation}
\begin{split}
    &\omega^2(\tau+b)+\omega(\omega_*-2\omega_D-6\omega_D(\tau+b))\\&+6\omega_D^2(1+\tau+b)+2\omega_D\omega_*(\eta-2)=0,
    \end{split}
    \label{dispITG}
\end{equation}
where $\tau=T_i/T_e$ is the ion to electron temperature ratio, $b=k_\perp^2\rho_i^2$ (which is assumed to be small), $\rho_i$ is the ion Larmor radius, $k_\perp$ is the perpendicular wavenumber, $\eta=L_{ni}/L_{Ti}$, where $L_{ni}=-d(\ln{n_i})/dx$ is the ion density logarithmic gradient, $L_{Ti}=-d(\ln{T_i})/dx$ is the ion temperature logarithmic gradient, $\omega_D=\textbf{v}_D\cdot\textbf{k}_\perp$ is the magnetic drift frequency and $\omega_*=\textbf{v}_*\cdot\textbf{k}_\perp$ is the diamagnetic drift frequency. Equation \eqref{dispITG} can be solved to find
\begin{widetext}
\begin{equation}
    \omega=\frac{2\omega_D+6\omega_D(\tau+b)-\omega_*}{2(\tau+b)}\pm\frac{\sqrt{(\omega_*-2\omega_D-6\omega_D(\tau+b))^2-4(\tau+b)(6\omega_D^2(1+\tau+b)+2\omega_D\omega_*(\eta-2))}}{2(\tau+b)}.
    \end{equation}
    \end{widetext}
We can focus on the second term, as it corresponds to the growth rate $\gamma$ of the mode. Moreover, to further simplify the expression, we will consider the flat density limit, which corresponds to $\eta\rightarrow\infty$ and $\omega_*\rightarrow0$. Therefore, we can write
\begin{equation}
\gamma=\pm\frac{\sqrt{2(\tau+b)\omega_D\omega_*\eta-3\omega_D^2(\tau+b)^2-\omega_D^2}}{(\tau+b)}.
\end{equation}
Therefore, the condition for which the growth rate of the mode $\gamma$ is maximized is given by
\begin{equation}
    \omega_D=\frac{(\tau+b)\eta}{1+3(\tau+b)^2}\omega_*.
\end{equation}
This is the optimal condition for the onset of an unstable toroidal ITG.

\section{Field-aligned coordinates and GENE}\label{B}

GENE, like many other GK codes, uses a field-aligned coordinate system to discretize the domain. This choice is made to exploit the large anisotropy of turbulence in order to reduce the computational cost. The field-aligned coordinate system is made of three coordinates: $x$ selects a flux surface, $y$ selects a field line on the flux surface and $z$ selects a point on the field line. The magnetic field can be written in Clebsch form as
\begin{equation}
    \textbf{B}=\boldsymbol{\nabla}x\wedge\boldsymbol{\nabla}y.
    \label{Clebsh}
\end{equation}
It is clear that this is not an orthogonal coordinate system, but a curvilinear one. Hence, the standard algebra for curvilinear systems can be applied \cite{Flux_coord}. We can define the metric coefficients $g^{ij}$ and $g_{ij}$, which define the differential arc length along a curve in the curvilinear system
\begin{equation}
    g_{ij}=\frac{\partial \textbf{R}}{\partial u^i}\cdot\frac{\partial \textbf{R}}{\partial u^j}
\end{equation}
\begin{equation}
    g^{ij}=\boldsymbol{\nabla}u^i\cdot\boldsymbol{\nabla}u^j
\end{equation}
where \textbf{R} is the position in a Cartesian coordinate system and the $u^i$ are the curvilinear coordinates. Hence, using GENE notation, $u^1=x$, $u^2=y$, $u^3=z$. 

We can also find the Jacobian of the transformation from Cartesian to curvilinear coordinates
\begin{equation}
    \mathcal{J}=\frac{\partial \textbf{R}}{\partial u^1}\cdot\left[\frac{\partial \textbf{R}}{\partial u^2}\wedge\frac{\partial \textbf{R}}{\partial u^3}\right],
\end{equation}
which is related to the metric coefficients through \cite{Flux_coord}
\begin{equation}
    \sqrt{\det{[g^{ij}]}}=\mathcal{J}.
\end{equation}

\section{Derivation of magnetic drift frequency geometric dependence}\label{C}
In this appendix we show how equation \eqref{omegaD_OK} can be derived.The magnetic drift frequency $\omega_D$ can be written as
\begin{equation}
\begin{split}
    \omega_D&=\textbf{v}_D\cdot\textbf{k}_\perp\\&=\left[\frac{1}{B\Omega}\left(\frac{v_\perp^2}{2}+v_\parallel^2\right)\textbf{b}\wedge\boldsymbol{\nabla}B\right]\cdot\left[ k_x\boldsymbol{\nabla}x+k_y\boldsymbol{\nabla}y\right]\\&=V\left[\textbf{b}\wedge\boldsymbol{\nabla}B\right]\cdot\left[ k_x\boldsymbol{\nabla}x+k_y\boldsymbol{\nabla}y\right],
    \end{split}
    \label{omegaD}
\end{equation}
where we used the definitions of $\textbf{v}_D$ and $\textbf{k}_\perp$, and we have hidden all the constants in $V$.  Now, to make all the terms explicit, we can use the mixed product rule in curvilinear coordinates
\begin{equation}
    (\textbf{A}\wedge\textbf{B})\cdot\textbf{C}=\mathcal{J}\epsilon_{ijk}A^jB^kC^i,
    \label{mixed}
\end{equation}
where $\mathcal{J}$ is the Jacobian and we made use of Einstein's notation for repeated indices. As will be clear in the following, it is convenient to use the covariant version of $\textbf{B}$ and $\textbf{C}$. Hence, using the contravariant metric tensor $g^{ij}$, we can write equation \eqref{mixed} as
\begin{equation}
    (\textbf{A}\wedge\textbf{B})\cdot\textbf{C}=\mathcal{J}\epsilon_{ijk}g^{km}g^{il}A^jB_mC_l.
    \label{mixed1}
\end{equation}
Then, we can set $\textbf{A}=\textbf{b}$ and use the fact that $\textbf{b}\parallel\textbf{z}$ (the coordinate parallel to the magnetic field lines). Hence, $b^j=b\delta^{j3}$, where $\delta^{ij}$ is Kronecker's delta. Therefore, we can simplify equation \eqref{mixed1}
\begin{equation}
    (\textbf{b}\wedge\textbf{B})\cdot\textbf{C}=\mathcal{J}b\epsilon_{i3k}g^{km}g^{il}B_mC_l,
\end{equation}
which gives
\begin{equation}
    (\textbf{b}\wedge\textbf{B})\cdot\textbf{C}=\mathcal{J}b\left(\epsilon_{132}g^{2m}g^{1l}B_mC_l+\epsilon_{231}g^{1m}g^{2l}B_mC_l\right).
\end{equation}
Hence, by applying the Levi-Civita tensor rule
\begin{equation}
    (\textbf{b}\wedge\textbf{B})\cdot\textbf{C}=\mathcal{J}b\left(g^{1m}g^{2l}-g^{2m}g^{1l}\right)B_mC_l.
    \label{mixed2}
\end{equation}
Now, since the metric tensor $g$ is symmetric, all the terms with $l=m$ will cancel each other. Therefore, by expanding  equation \eqref{mixed2}, we can write
\begin{equation}
    \begin{split}
        (\textbf{b}\wedge\textbf{B})\cdot\textbf{C}=&\mathcal{J}b[(g^{11}g^{22}-g^{21}g^{12})B_1C_2\\&+(g^{11}g^{23}-g^{21}g^{13})B_1C_2\\&+(g^{12}g^{21}-g^{22}g^{11})B_2C_1\\&+(g^{12}g^{23}-g^{22}g^{13})B_2C_3\\&+(g^{13}g^{21}-g^{23}g^{11})B_3C_1\\&+(g^{13}g^{22}-g^{23}g^{12})B_3C_2].
    \end{split}
    \label{mixed3}
\end{equation}
We can define some useful quantities to simplify the previous equation
\begin{align}
    \gamma_1=g^{11}g^{22}-g^{21}g^{12} \\
    \gamma_2=g^{11}g^{23}-g^{21}g^{13},\\
    \gamma_3=g^{12}g^{23}-g^{22}g^{13}
\end{align}
which allow us to write equation \eqref{mixed3} as
\begin{equation}
    \begin{split}
    (\textbf{b}\wedge\textbf{B})\cdot\textbf{C}&=\mathcal{J}b[\gamma_1(B_1C_2-B_2C_1)+(\gamma_2B_1+\gamma_3B_2)C_3\\&-B_3(\gamma_2C_1+\gamma_3C_2)].
    \end{split}
\end{equation}
Finally, if we replace $\textbf{B}=\boldsymbol{\nabla}B$ and $\textbf{C}=k_x\boldsymbol{\nabla}x+k_y\boldsymbol{\nabla}y$ in the previous equation, we can write equation \eqref{omegaD} as   
\begin{equation}
    \omega_D=\Lambda\left(k_x\underbrace{\left(-\partial_y B-\frac{\gamma_2}{\gamma_1}\partial_z B\right)}_{v_{Dx}}+k_y\underbrace{\left(\partial_x B -\frac{\gamma_3}{\gamma_1}\partial_z B\right)}_{v_{Dy}}\right)
    \label{omegaD_ok}
\end{equation}
where the scalar $\Lambda$ comprises all the constants, $v_{Dx}$ is the radial component of the magnetic drift velocity and $v_{Dy}$ the binormal one.

\section{NT ITG physical picture applied to elongation and magnetic shear scans}\label{D}

In this appendix, we apply our NT ITG physical picture to elongation and magnetic shear scans performed for the large $A$ pITG-2 scenario with adiabatic electrons. While in general it is necessary to use kinetic electrons to accurately model NT, previous simulations showed that adiabatic electrons are sufficient in this particular case.

\subsection{Elongation scan}

\begin{figure}[b]
    \centering
    \includegraphics[width=\linewidth]{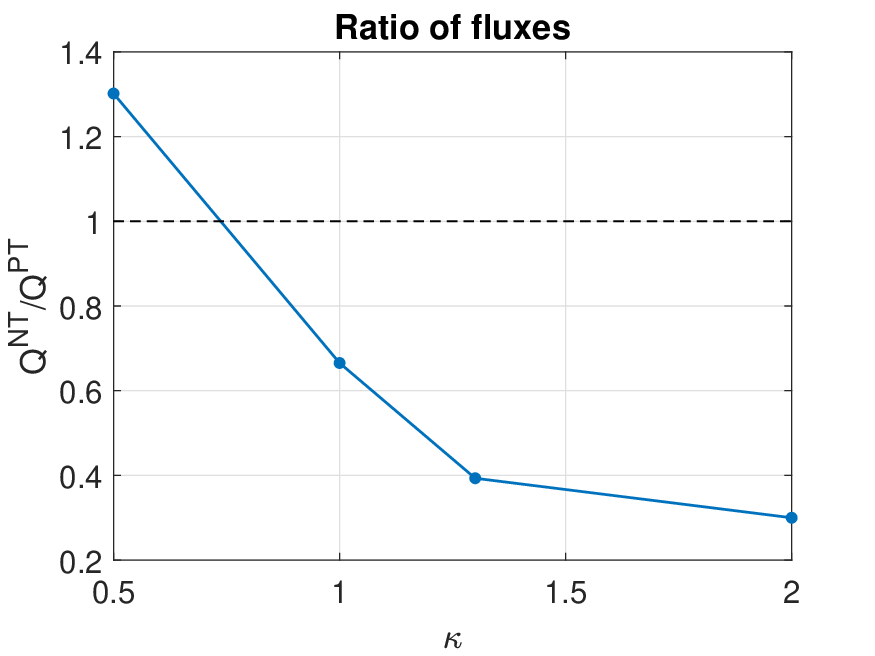}
    \caption{Ratio of NT heat flux over heat flux of the PT counterpart as a function of the plasma elongation. The dashed black line highlight a ratio of 1, where NT and PT have the same confinement.}
    \label{Qratio_kappa}
\end{figure}

We took the NT and PT large $A$ pITG-2 equilibria and performed NL simulations where the elongation was changed in a range from $0.5$ to $2.0$. Figure \ref{Qratio_kappa} shows the ratio between the nonlinear heat fluxes from the NT cases and the PT cases as a function of the elongation. We observe a monotonic decrease in the ratio with elongation. Therefore, larger values of $\kappa$ increase the beneficial effect of NT on confinement and very low values of elongation make NT detrimental. 

\begin{figure}[]
    \centering
    \begin{subfigure}[Binormal component of magnetic drift velocity]
        { \includegraphics[width=\linewidth]{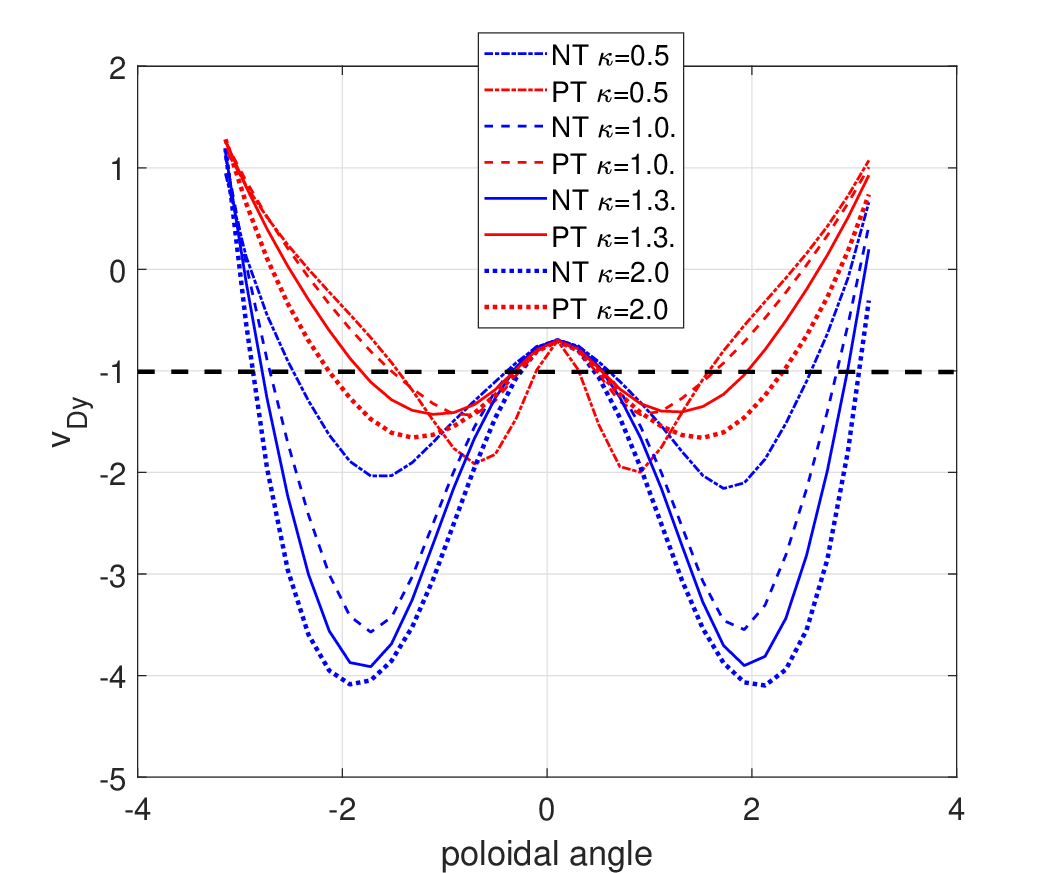}}
    \end{subfigure}
    \begin{subfigure}[$g^{yy}$]
        { \includegraphics[width=\linewidth]{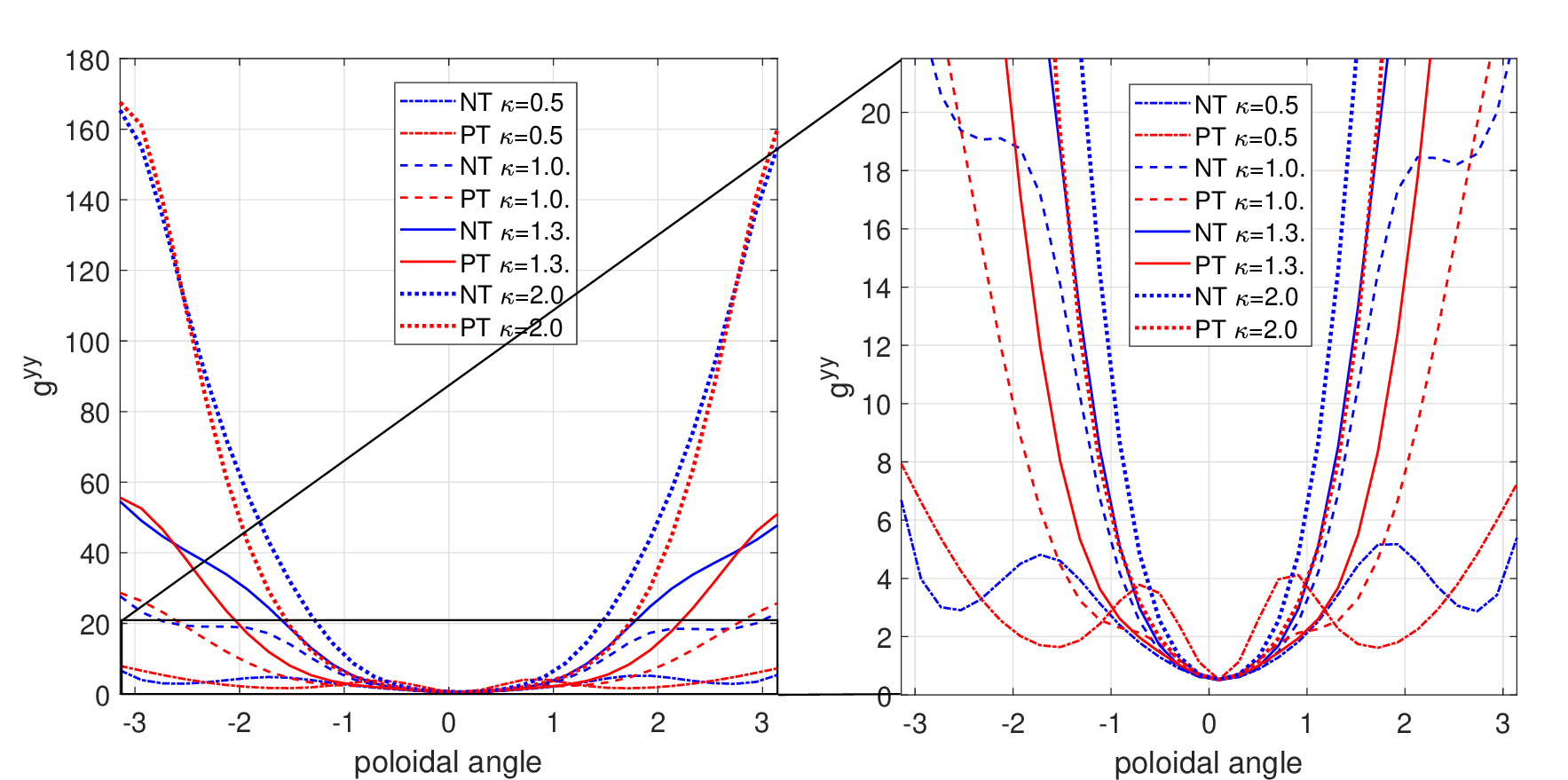}}
    \end{subfigure}
    \caption{(a) Binormal component of the magnetic drift velocity and (b) $g^{yy}$ component of the metric tensor as a function of the poloidal angle, for different values of elongation and opposite signs of triangularity.}
    \label{vDy_kappa}
\end{figure}
To understand this behaviour, we can apply the NT ITG physical picture and explain the trend in terms of magnetic drifts and FLR effects. Figure \ref{vDy_kappa} shows how the binormal component of the magnetic drifts $v_{Dy}$ and the $g^{yy}$ element of the metric tensor change as the elongation is varied. We observe that the behaviour of these two factors correlates well with the beneficial effect of NT. Indeed, as the elongation is decreased, the drift velocity of NT geometry varies less with poloidal angle. For $\kappa=0.5$ it is actually closer to the resonance condition than PT around the outboard midplane. Similar behaviour can be observed for the FLR effect. At large elongation, the $g^{yy}$ metric of the NT geometry is above the one of the PT case. However, when elongation is decreased below $\kappa=1$, NT's $g^{yy}$ is below PT's around the outboard midplane. Therefore, applying our NT ITG physical picture, we can say that the situation is the opposite when at $\kappa=0.5$ compared to the rest of the paper: it is PT that has faster magnetic drifts and stronger FLR effects.

\subsection{Magnetic shear scan}

\begin{figure}[]
    \centering
    \includegraphics[width=\linewidth]{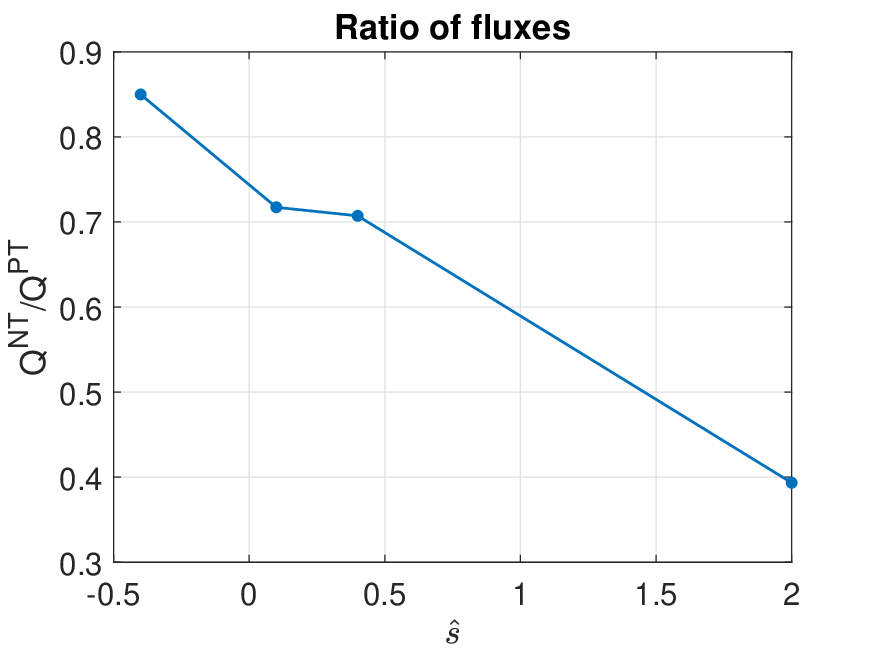}
    \caption{Ratio of NT heat flux over heat flux of the PT counterpart as a function of the plasma magnetic shear.}
    \label{Qratio_shear}
\end{figure}

We performed the same study as in the previous section, but with the magnetic shear $\hat{s}$ instead of elongation. Magnetic shear $\hat{s}$ was changed from large positive to negative values of shear, in the range $\hat{s}=[-0.5,2.0]$. Figure \ref{Qratio_shear} shows the ratio between the nonlinear heat flux from the NT cases and the PT cases as a function of the magnetic shear. We observe a monotonic decrease of the ratio with $\hat{s}$.  We could not find a value of magnetic shear for which NT becomes detrimental, but NT becomes less beneficial at lower $\hat{s}$. These observations are consistent with \cite{merlo_jenko_2023}. However, here we can try to apply our NT ITG physical picture to explain this trend.

\begin{figure}[]
    \centering
    \begin{subfigure}[Binormal component of magnetic drift velocity]
        { \includegraphics[width=\linewidth]{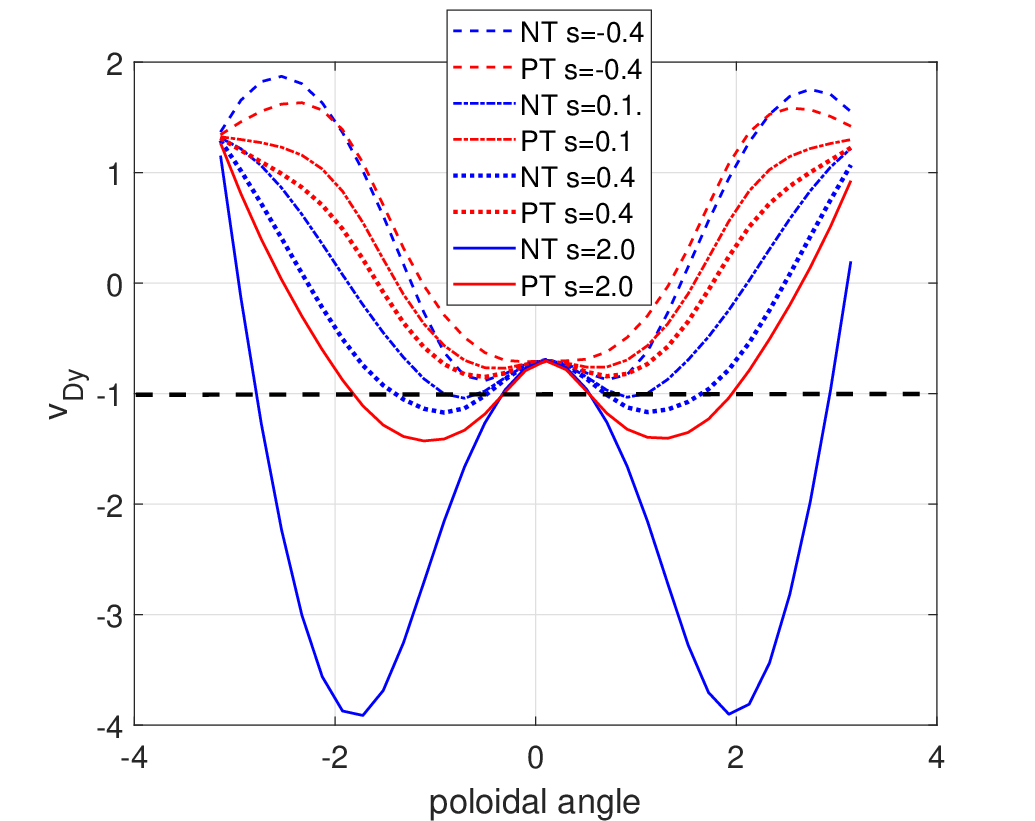}}
    \end{subfigure}
    \begin{subfigure}[$g^{yy}$]
        { \includegraphics[width=\linewidth]{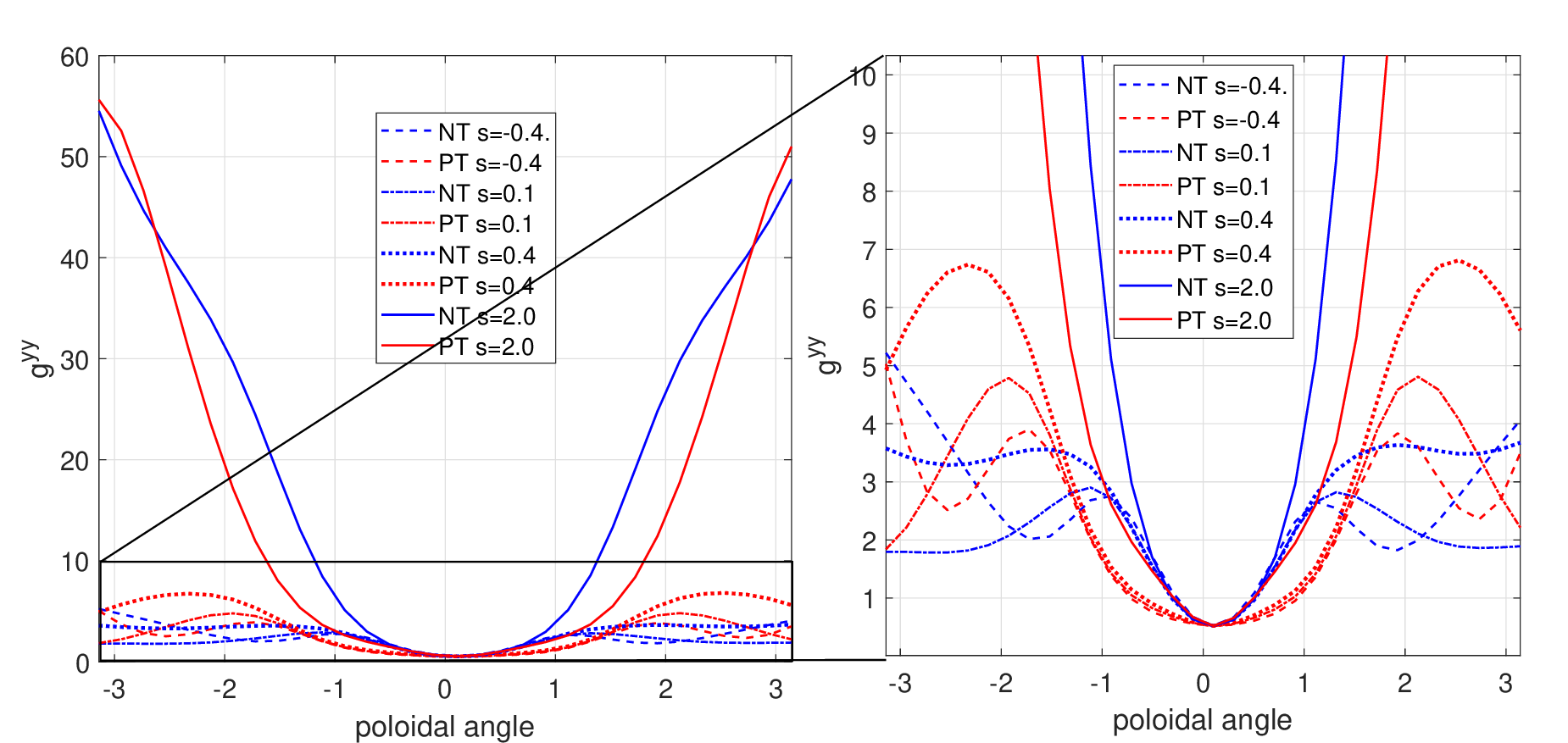}}
    \end{subfigure}
    \caption{(a) Binormal component of the magnetic drift velocity and (b) $g^{yy}$ component of the metric tensor as a function of the poloidal angle, for different values of magnetic shear and opposite signs of triangularity.}
    \label{vDy_shear}
\end{figure}

In figure \ref{vDy_shear} we show the binormal component of the magnetic drift $v_{Dy}$ and the $g^{yy}$ metric element as a function of the poloidal angle, for different values of magnetic shear. We observe that as the magnetic shear is decreased towards negative values, $v_{Dy}$  becomes more similar between the two geometries, though NT remains slightly farther away from the resonance than PT. If we look at $g^{yy}$, we observe that the same is true for FLR effects - they become more similar as $\hat{s}$ is reduced, though FLR effects remain slightly stronger in NT at the lowest value of $\hat{s}$. Therefore, we can conclude that the beneficial effect of NT on ITG decreases at low magnetic shear because the drift velocity of ions gets closer to the resonance condition, and the FLR effects are reduced.

\bibliography{refs}

\end{document}